\documentclass[conference,compsoc]{IEEEtran}

\usepackage{hyperref}

\usepackage{amsmath}
\usepackage{filecontents}
\usepackage{xspace}
\usepackage[linesnumbered,ruled,vlined]{algorithm2e}
\usepackage{amssymb}
\usepackage{enumitem}
\usepackage{pifont}
\usepackage{colortbl}
\usepackage{xcolor}
\usepackage{graphicx}
\usepackage{booktabs}
\usepackage{fontawesome}
\usepackage{transparent}
\usepackage{titlesec}
\usepackage{titlesec}
\usepackage{multirow}
\usepackage{amsfonts}
\usepackage{mathtools}

\DeclareMathAlphabet{\mathcal}{OMS}{cmsy}{m}{n}

\usepackage{caption}
\definecolor{LightCyan}{rgb}{0.88,1,1}
\definecolor{lightgreen}{HTML}{90ee90}
\definecolor{lightblue}{HTML}{badde8}
\newcommand{\sys}{{\sc BandFuzz}\xspace}
\newcommand{\autofz}{{autofz}\xspace}

\def\Snospace~{\S{}}

\begin{document}
\makeatletter
\newcommand{\linebreakand}{%
  \end{@IEEEauthorhalign}
  \hfill\mbox{}\par
  \mbox{}\hfill\begin{@IEEEauthorhalign}
}
\makeatother

\title{BandFuzz: An ML-powered Collaborative Fuzzing Framework}
\author{
\IEEEauthorblockN{Wenxuan Shi}
\IEEEauthorblockA{Northwestern University\\
wenxuan.shi@northwestern.edu}
\and
\IEEEauthorblockN{Hongwei Li}
\IEEEauthorblockA{UC Santa Barbara\\
hongwei@ucsb.edu}
\and
\IEEEauthorblockN{Jiahao Yu}
\IEEEauthorblockA{Northwestern University\\
jiahao.yu@northwestern.edu}
\and
\IEEEauthorblockN{Xinqian Sun}
\IEEEauthorblockA{Northwestern University\\
xinqiansun2027@u.northwestern.edu}
\linebreakand
\IEEEauthorblockN{Wenbo Guo}
\IEEEauthorblockA{UC Santa Barbara\\
henrygwb@ucsb.edu}
\and
\IEEEauthorblockN{Xinyu Xing}
\IEEEauthorblockA{Northwestern University\\
xinyu.xing@northwestern.edu}
}

\maketitle
\begin{abstract}

Collaborative fuzzing has recently emerged as a technique that combines multiple individual fuzzers and dynamically chooses the appropriate combinations suited for different programs. 
Unlike individual fuzzers, which rely on specific assumptions to maintain their effectiveness, collaborative fuzzing relaxes the assumptions on target programs, providing constant and robust performance across various programs.
Ideally, collaborative fuzzing should be a more promising direction toward generic fuzzing solutions, as it mitigates the need for manual cherry-picking of individual fuzzers.
However, the effectiveness of existing collaborative fuzzing frameworks is limited by major challenges, such as the need for additional computational resources compared to individual fuzzers and the inefficient allocation of resources among the various fuzzers.

To tackle these challenges, we present \sys, an ML-powered collaborative fuzzing framework that outperforms individual fuzzers without requiring additional computational resources.
The key technical contribution of \sys lies in its novel resource allocation algorithm driven by our proposed multi-armed bandits model.
Different from the greedy methods employed in existing collaborative fuzzing frameworks, \sys models the long-term impact of individual fuzzers, enabling the discovery of globally optimal collaborative strategies.
Moreover, we propose a novel fuzzer evaluation method that accesses not only code coverage but also the fuzzer's capability of solving difficult branches. 
Finally, we integrate a real-time seed synchronization mechanism, as well as a set of implementation-wise optimizations to improve fuzzing efficiency and stability.
Through extensive experiments on Fuzzbench and Fuzzer Test Suite, we show that \sys outperforms state-of-the-art collaborative fuzzing framework, \autofz, and widely used individual fuzzers. 
We also verify \sys's key designs through a comprehensive ablation study.
Notably, we show \sys's effectiveness in real-world bug detection by analyzing the results of a worldwide fuzzing competition, where \sys won the first place. 

\end{abstract}
\section{Introduction}
\label{sec:introduction}

The main purpose of fuzz testing is to thoroughly explore a program's state space, uncovering potential vulnerabilities hidden in the program states. 
A comprehensive fuzzing test ensures the program is carefully tested with a wide variety of inputs, seeking to uncover as many vulnerabilities as possible.
In recent years, there has been a significant research emphasis on expanding the program states that fuzzing can cover through optimizing key fuzzing components. 
These improvements include better techniques for selecting seeds~\cite{she2019neuzz,she2020mtfuzz,li2019cerebro,wang2020not,zhao2022alphuzz,wang2021reinforcement,rebert2014optimizing,liang2019deepfuzzer,herrera2021seed,gan2018collafl,zhu2023better}, more efficient power scheduling~\cite{bohme2016CoveragebasedGreyboxFuzzing,yue2020ecofuzz}, and more effective ways to mutate the inputs~\cite{rawat2017vuzzer,lemieux2018FairFuzzTargetedMutation,jauernig2023DARWINSurvivalFittest,lyu2022MOPTOptimizedMutation,aschermann2019REDQUEENFuzzingInputtoState,2023AkiHelinRadamsa,you2019profuzzer,wang2021cmfuzz,lee2023learning}. 
All of these help fuzz testing cover more code branches for the program under test.

However, recent studies have shown progress in fuzzing technologies is slowing down~\cite{schloegel2024sok}. 
The latest public benchmarks~\cite{2023fuzzbenchreport} and results from open competitions~\cite{liu2023sbft} support this. 
They show a lack of new academic fuzzers that can outperform well-established industry tools like AFL++~\cite{fioraldi2020afl++} from 2020. 
Reports from FuzzBench~\cite{2024fuzzbenchreport} and contests like 2023 SBFT Fuzzing Competition~\cite{2023sbft} also highlight that few fuzzers consistently perform better than others across different test targets. 
This shortfall mainly comes from the reliance on specific heuristics and assumptions in the design of research fuzzers.
These assumptions may not match up well with the complexities seen in real-world programs, resulting in significant disparities in fuzzing performance.

Recognizing the limitations of individual fuzzing tools, recent research has started looking into collaborative fuzzing. 
This approach combines the strengths of multiple fuzzers into one unified fuzzer. 
The reasoning behind collaborative fuzzing is straightforward yet compelling: different fuzzers are capable of exploring different programs and program states. 
By combining the unique capabilities of various fuzzers, it is more likely to explore more comprehensive program states, achieving better coverage across diverse programs. 

Recent studies have shown that collaborative fuzzers often outperform individual ones (e.g.,~\cite{osterlund2021collabfuzz,guler2020cupid,chen2019enfuzz,fu2023autofz}).
However, their effectiveness greatly relies on using a lot more additional computational resources compared to individual fuzzers. 
For example, recently collaborative fuzzing tools like EnFuzz~\cite{chen2019enfuzz}, CollabFuzz~\cite{osterlund2021collabfuzz}, and Cupid~\cite{guler2020cupid} all need to run multiple fuzzers in parallel at the same time. 
This greatly hinders the usage of collaborative fuzzers, especially when fuzzing is done on systems with limited computational resources. 
As such, it is critical for collaborative fuzzers to efficiently~\textit{allocate resources} among the individual fuzzers they employ. 
This involves addressing the following three main challenges.
First, to establish an effective way for the fuzzers to share knowledge. 
In a collaborative setting, different fuzzers may repeatedly explore the same code branches. 
Without a mechanism for the fuzzers to exchange code coverage information, these overlapping efforts could waste a lot of computing resources, undermining the overall fuzzing efficiency.
%
Second, to develop a real-time fuzzer evaluation method that can accurately and dynamically evaluate the performance of each fuzzer during the fuzzing process. 
Effective resource allocation depends on identifying and prioritizing the most effective fuzzers. 
Without an accurate and real-time fuzzer evaluation, determining which fuzzers to assign more resources becomes difficult.
%
Third, to implement a globally optimal resource allocation strategy that prioritizes not only the current effectiveness of individual fuzzers, but more importantly, their long-term impact.
As a counter-example, we show that the greedy approach utilized by a recent collaborative fuzzer \autofz~\cite{fu2023autofz} is sub-optimal, resulting in limited performance and potentially increasing the waste of computational resources.

To tackle these challenges, we introduce \sys, an innovative collaborative fuzzing framework. 
Our key technical novelty is a multi-arm bandits-powered resource allocation method.
Unlike greedy methods that allocate computational resources solely according to the immediate performance of individual fuzzers, our bandits-based approach balances fuzzers' long-term effectiveness with their current performance, resulting in a more globally optimal resource allocation strategy.
Furthermore, benefiting from the parametric efficiency of multi-arm bandits, our resource allocation method introduces only minimal computational overhead and does not require large-scale samples to learn an effective strategy.
As explained in~\autoref{subsec:tech_allocate}, standard bandit algorithms cannot be directly applied to our problem. 
As such, we introduce customized designs to a commonly used bandit algorithm, Thompson sampling, to make it suitable for our specific needs.

Our framework also integrates an innovative method for evaluating fuzzers. 
Our approach goes beyond the simple branch coverage metric and considers the effort needed to explore new branches. 
Specifically, we introduce a new metric to quantify the difficulty of exploring new branches and integrate it with coverage to form our final fuzzer evaluation metric. 
As shown in~\autoref{sec:eval}, our proposed metric provides a better assessment of a fuzzer's real-world performance.
Furthermore, we develop a real-time seed synchronization mechanism to prevent individual fuzzers from redundantly exploring the same branches. 
We also conduct non-trivial implementation optimizations to enable real-time fuzzer evaluation and enhance the stability of our framework.

Through extensive evaluation on various fuzzing targets from two benchmarks, FuzzBench~\cite{metzman2021fuzzbench} and FTS~\cite{fuzzer_test_suite_2021}, we first show that \sys outperforms ten popular fuzzers (e.g., AFL++~\cite{fioraldi2020afl++}, AFL~\cite{zalewski2020afl}, Angora~\cite{chen2018angora}, Darwin~\cite{jauernig2023DARWINSurvivalFittest}, MOPT~\cite{lyu2022MOPTOptimizedMutation}
) integrated in our framework. 
Second, we demonstrate that \sys significantly outperforms the state-of-the-art collaborative fuzzing technique, autofz~\cite{fu2023autofz}, confirming the superiority of our bandits-based resource allocations over greedy methods. 
Third, we conduct a comprehensive ablation study to verify the necessity of our three key components: seed synchronization, fuzzer evaluation, and bandits-based resource allocation.
Fourth, we run a sensitivity test to confirm the effectiveness of our customized bandits model and the in-sensitivity to key hyper-parameters.  
More importantly, we used \sys to participate in a recent world-level fuzzing competition.
\sys won the~\textit{first place}, beating state-of-the-art research and industrial-level fuzzers. 
The competition also highlights \sys's superior capability in vulnerability detection. 
To the best of our knowledge, \sys is the first ML-powered collaborative fuzzing framework that demonstrates its effectiveness and stability across a wide range of programs.

In summary, we make the following contributions.

\begin{itemize}[leftmargin=*]

    \item We introduce \sys, an ML-powered collaborative fuzzing framework. \sys integrates an accurate and real-time fuzzer evaluation method, a real-time seed synchronization method, and a novel mechanism that uses our customized bandits algorithm to allocate resources.

    \item We implement \sys to incorporate ten widely-used fuzzers, demonstrating its superiority over state-of-the-art individual and collaborative fuzzing tools.

    \item We comprehensively evaluate the necessity of three main designs in \sys and its insensitivity to key hyper-parameters.
    We further provide an analysis of a recent fuzzing competition result, where \sys beats cutting-edge research fuzzers and industrial-level fuzzers. 
\end{itemize}

\section{Background}
\label{sec:bg}

A wide range of research has been conducted on program fuzzing, encompassing everything from the development of individual fuzzers to the implementation of collaborative fuzzing that combines multiple fuzzers.
In the following, we briefly summarize these research endeavors, with a primary focus on recently emerged collaborative fuzzing frameworks and their limitations.

\subsection{Individual Fuzzers}
\label{subsec:bg_fuzzer}

Existing research on individual fuzzer development mainly focuses on improving the following components: seed selection, power scheduling, mutation strategies, and integration of complex analysis methods.

\noindent{\bf Seed selection. }
Beyond standard seed selection strategies that consider execution efficiency and seed length, research in this direction has explored other factors~\cite{wang2020not,zhao2022alphuzz,wang2021reinforcement,rebert2014optimizing,liang2019deepfuzzer,herrera2021seed,gan2018collafl,zhu2023better,bohme2016CoveragebasedGreyboxFuzzing, rawat2017vuzzer, zhao2019send,she2020mtfuzz}.
For example, NEUZZ~\cite{she2019neuzz} considers branch uniqueness, Cerebro~\cite{li2019cerebro} leverages code complexity, and EcoFuzz~\cite{yue2020ecofuzz} uses state transition probability.

\noindent{\bf Power scheduling. }
Another line of research focuses on improving power scheduling~\cite{bohme2016CoveragebasedGreyboxFuzzing,yue2020ecofuzz,liang2019deepfuzzer,pham2019smart,zhang2022path,chen2022novel}.
For example, AFLFast~\cite{bohme2016CoveragebasedGreyboxFuzzing} leverages a Markov chain model to allocate more energy to seeds that reach paths that are infrequently visited.

\noindent{\bf Mutation strategies. } 
This research area focuses on creating new mutation techniques or mutator scheduling strategies~\cite{rawat2017vuzzer,lemieux2018FairFuzzTargetedMutation,jauernig2023DARWINSurvivalFittest,lyu2022MOPTOptimizedMutation,aschermann2019REDQUEENFuzzingInputtoState,2023AkiHelinRadamsa,you2019profuzzer,wang2021cmfuzz,lee2023learning}. 
Notably, FairFuzz~\cite{lemieux2018FairFuzzTargetedMutation} prioritize the mutation of input bytes linked to rarely executed branches. 
RedQueen~\cite{aschermann2019REDQUEENFuzzingInputtoState} develops another method for selecting important input bytes to prioritize mutations. 
DARWIN~\cite{jauernig2023DARWINSurvivalFittest} and MOpt~\cite{lyu2022MOPTOptimizedMutation} design a novel mutator scheduling strategy based on the evolution strategy and Particle Swarm Optimization (PSO), respectively. 

\noindent{\bf Integration of concolic execution and taint analysis. } 
Some fuzzers incorporate concolic execution~\cite{stephens2016driller,chen22SYMSANTimeSpace,poeplau2020symbolic,yun2018qsym,poeplau2021symqemu,huang2020pangolin,borzacchiello2021fuzzolic,shen2022drifuzz} or taint analysis~\cite{chen2018angora,liang2022pata,gan2020greyone,kim2020hfl} to better tackle difficult branches and reduce fuzzing search space, enhancing the overall fuzzing efficiency. 
Some representative fuzzers include Angora~\cite{chen2018angora}, SymCC~\cite{poeplau2020symbolic}, and SymSan~\cite{chen22SYMSANTimeSpace}, which leverage these advanced techniques to explore challenging code paths more effectively.

\noindent{\bf Limitations.}
Although individual fuzzers exhibit state-of-the-art performance in specific contexts, their effectiveness may decrease under different conditions, resulting in large variations in fuzzing performances across different programs.  
This variation is due to the specific assumptions that each fuzzer relies on, which may not apply to all programs. 
Consequently, no single fuzzer emerges as a universally optimal solution, capable of delivering consistently superior performance across a wide range of programs.

\subsection{Collaborative Fuzzing}
\label{subsec:bg_rw}

To tackle the limitation of individual fuzzers, recent research is shifting focus to collaborative fuzzing that integrates the strengths of multiple fuzzers to enhance the fuzzing effectiveness and stability across diverse programs. 

\noindent{\bf EnFuzz}~\cite{chen2019enfuzz} integrates six fuzzers, which are handpicked by domain experts. 
EnFuzz runs the selected fuzzers in parallel, each taking a separate CPU core. 
It also uses a dedicated thread to synchronize seeds every two minutes (i.e.,  get a union seed set and copy it to every fuzzer), ensuring a shared and up-to-date seed set for all participants.

\noindent{\bf CollabFuzz}~\cite{osterlund2021collabfuzz} retains EnFuzz's architecture and fuzzers but extends its capabilities with advanced seed synchronization policies, i.e., providing more options for what seeds to synchronize (e.g., prioritizing important seeds) and when to synchronize (e.g., immediately upon a seed is generated).

\noindent{\bf Cupid}~\cite{guler2020cupid} 
first conducts an offline evaluation of eight fuzzers on some pre-selected programs and records the probability for each fuzzer to cover each branch.
It requires the user to decide the number of fuzzers to use. 
Then, Cupid selects the optimal fuzzer combination that maximizes the average branch coverage.
Like EnFuzz and CollabFuzz, each fuzzer runs on a separate CPU core, with seed synchronization akin to EnFuzz's approach.

\noindent{\bf \autofz}~\cite{fu2023autofz} is different from the approaches of EnFuzz, CollabFuzz, and Cupid. 
It is tailored to perform fuzzing under resource constraints. 
Specifically, it proposes a straightforward exploration-exploit method for resource allocation among the fuzzers it integrates. 
During the exploration phase, \autofz allocates the same resource to each fuzzer and evaluates them by unique coverage.
Based on these evaluations, the subsequent exploitation phase allocates more resources to fuzzers that demonstrated superior performance during the exploration phase. 
This allocation strategy essentially employs a greedy algorithm that uses locally optimal solutions in each fuzzing period to approximate a globally optimal solution in the entire fuzzing campaign.

\begin{figure}[t]
    \centering
    \includegraphics[width=1\linewidth]{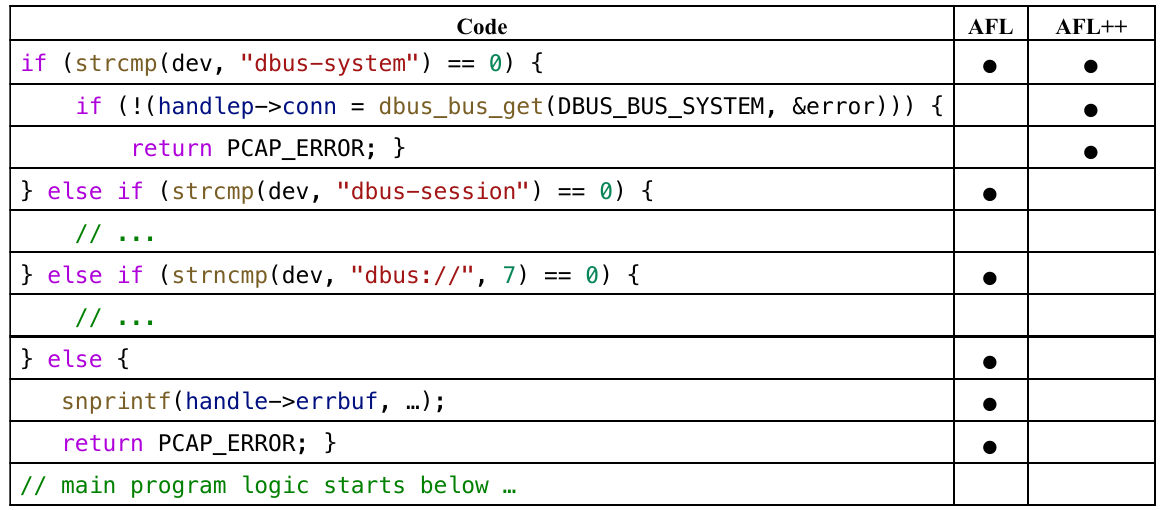}
    \caption{An example of two individual fuzzer's code coverage on \texttt{libpcap}. The dense cycle points out the resolved branches.}
    \label{fig:coverage_example}
    \vspace{-2mm}
\end{figure}

\noindent{\bf Limitations.}
Despite the novel designs, existing collaborative fuzzing techniques have not achieved the same level of efficacy as widely used individual fuzzers. 
Below, we outline the reasons for this discrepancy.

\begin{itemize}[leftmargin=*]

    \item \noindent{\bf Requiring additional computational resources.} 
    EnFuzz, CollabFuzz, and Cupid run selected fuzzers in parallel, which means they all necessitate more computational resources than individual fuzzers. 
    Consequently, enhancements in fuzzing performance can be attributed to the increased resources used. This design limits their effectiveness when resources are constrained.
    
    \item \noindent{\bf Infrequent cross-fuzzer knowledge/seed synchronization.} 
    During fuzz testing, fuzzers included in \autofz and other collaborative fuzzing frameworks either do not share their path coverage or do so infrequently. 
    This results in a considerable waste of resources, as areas of the program space left unexplored by one fuzzer might already have been covered by another.

    \item \noindent{\bf Non-real-time and inaccurate fuzzer evaluation.}
    Existing collaborative fuzzers are inefficient in measuring and tracking the real-time effectiveness of individual fuzzers throughout the fuzzing procedure.
    Specifically, EnFuzz and CollabFuzz directly run the selected fuzzers without any evaluation.
    Cupid involves an offline evaluation, where the measurement of individual fuzzers remains unchanged throughout the fuzzing process.
    \autofz assesses each fuzzer based on the uniqueness of code coverage it achieves during its exploration phase. 
    Specifically, fuzzers that uncover more unique paths are rated higher.
    However, this approach may limit the fuzzing exploration to only local areas of the codebase. 
    For instance, as illustrated in~\autoref{fig:coverage_example}, AFL might achieve a higher unique branch coverage than AFL++ but is limited to the input format validation phase, failing to reach the deeper and more critical part of the main program logic. 
    In contrast, AFL++ is capable of identifying critical branches with complex conditions (Line 2), enabling subsequent fuzzing processes to reach the main program logic and potentially achieve a higher branch coverage at a later stage.
    Furthermore, \autofz's fuzzing evaluation is not real-time.

    \item \noindent{\bf Inefficiencies in resource allocation strategy.}
    EnFuzz, CollabFuzz, and Cupid allocate a CPU core to each fuzzer without any resource allocations and thus cannot adapt to performance changes in individual fuzzers.
    \autofz explicitly separates the fuzzing process into multiple sub-periods and formulates a fixed resource allocation strategy for each period. 
    This greedy approach does not consider potential relationships between different periods, lacking a global assessment. 
    As demonstrated in existing works~\cite{lyu2022MOPTOptimizedMutation, bohme2017directed,lemieux2018perffuzz, kim2020maxafl} and our evaluation in \autoref{sec:eval}, a decision-making process without a global perspective results in a strategy that is not globally optimal.
    
\end{itemize}


 %
\section{Key Techniques}
\label{sec:methods}

To address the limitations above, we design and develop \sys. 
At a high level, we first propose a novel evaluation method for real-time and accurate measurement of dynamic performance changes of individual fuzzers. 
More importantly, we design a real-time resource allocation mechanism utilizing multi-arm bandits, which accounts for both local and global assessments.
In this section, we first provide an overview of our design rationale and then delve into the technical details.

\begin{figure}[t]
    \centering
    \includegraphics[width=1\linewidth]{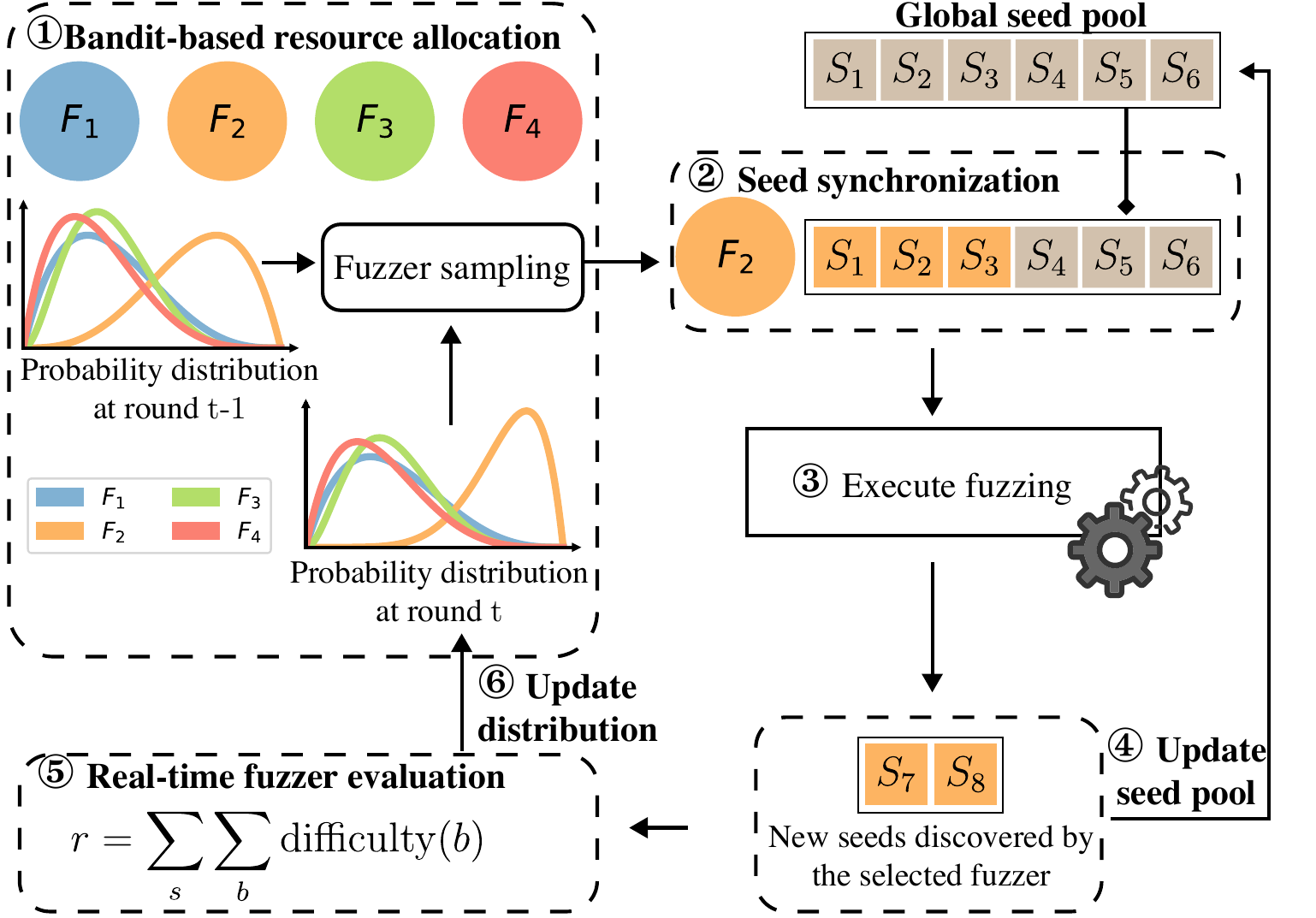}
    \caption{The overview of \sys. In the figure, \sys manages four fuzzers. In the current round $t$, the fuzzer $F_2$ is chosen. Initially, $F_2$'s seed pool contains seeds $S_1 \sim S_3$, while the global seed pool contains $S_1 \sim S_6$. After seed synchronization, $F_2$ extends its seed pool with the missing seeds and then proceeds to execute the fuzzing task, discovering new seeds $S_7$ and $S_8$. The new seeds are added to the global seed pool and evaluated to get a reward to update the probability distribution for $F_2$.}
    \label{fig:overview}
    \vspace{-2mm}
\end{figure}

\subsection{Overview of \sys}
\label{subsec:tech_overview}

\noindent{\bf Real-time and faithful fuzzing evaluation.}
As discussed in ~\autoref{subsec:bg_rw}, enabling real-time fuzzer evaluation is crucial to optimize resource allocation for more effective fuzzers.
To achieve this, a straightforward solution is to use each fuzzer's coverage at each round as the metric.
As detailed in~\autoref{sec:impl}, we conduct non-trivial implementation efforts to efficiently return the coverage of individual fuzzers at each fuzzing round.
However, coverage is not a proper metric.
First, different fuzzers may exhibit the same coverage, making it difficult to prioritize fuzzers with the same coverage.
Second, the difficulty and importance of identifying different branches varies a lot.  
As shown in~\autoref{fig:coverage_example}, a branch (Line 2) covered by AFL++ with a complex condition is more important than other branches, as it enables the exploration of main program logic.
Fuzzers unlocking more crucial branches should be deemed more effective than those addressing simpler ones. 
This evaluation cannot be achieved solely with coverage information.

To address these limitations, \autofz treats unique paths as difficult ones.
As discussed in~\autoref{subsec:bg_rw}, this metric may limit exploration to local regions.
Furthermore, it is computationally inefficient, making it difficult to conduct real-time evaluation of fuzzing dynamics. 
In this work, we propose a more efficient metric for measuring the branch difficulty (detailed in ~\autoref{subsec:tech_measure})
At a high level, we employ the time required to discover each branch as the measure of its difficulty. 
If a branch takes longer to be found, we consider it more challenging. 
Unlike branch uniqueness, recording branch discovery time introduces negligible overhead, significantly enhancing efficiency.
We integrate this metric with branch coverage and use the combined metric (denoted as~\textit{fuzzer evaluation metric}) to evaluate the effectiveness of individual fuzzers in each fuzzing round.

\noindent{\bf Multi-arm bandits based resource allocation.}
After enabling real-time evaluation of fuzzing dynamics, the next key step for effective collaborative fuzzing is to optimize resource allocation based on the effectiveness of individual fuzzers. 
Here, we discuss the single-core setup, and the goal is to assign more resources to more effective fuzzers throughout the fuzzing process.
Straightforward solutions involve either consistently selecting the best fuzzer after each evaluation or allocating a specific amount of resources to each fuzzer based on their effectiveness ranking, as employed in \autofz.
As discussed in~\autoref{subsec:bg_rw}, such greedy solutions lack the flexibility to explore allocation strategies that may not be optimal in the current period but could lead to better performance in longer runs.
As a result, such solutions fall short of achieving globally optimal performance throughout the entire fuzzing campaign.
Essentially, this is the problem of balancing exploitation and exploration.

We propose to leverage the multi-arm bandit to solve this problem.
Multi-arm bandits, or simply \textit{bandits}, represent a set of algorithms that learn efficient strategies for searching and scheduling problems~\cite{wolpert1997no,russo2018tutorial,thompson1933likelihood,zhang2022matching,auer2002finite}. 
Their objective is to strike a balance between exploiting the current optimal choices (exploitation) and exploring potentially better choices (exploration). 
Bandits naturally align with our problem of balancing exploitation and exploration in resource allocation. 
Furthermore, in contrast to other learning problems, bandits do not require pre-collection of a dataset or training a substantial number of parameters. 
Their real-time and efficient nature further sets them as a natural fit for our problem.
As we will detail in~\autoref{subsec:tech_allocate}, we build a customized bandit model for our problem, which treats individual fuzzers as an arm and our proposed fuzzer evaluation metric as the reward. 
We leverage the Thompson sampling algorithm to determine the weights for individual fuzzers in each fuzzing round and dynamically allocate resources in real time based on these weights.

\noindent{\bf Overall procedure of \sys.}
As demonstrated in~\autoref{fig:overview}, in each round, given a weight distribution of individual fuzzers, \sys first sample a fuzzer $F_2$ based on this distribution (i.e., $F_2$ is selected in round $t$) (\ding{192}).
It then conducts a seed synchronization (\ding{193}), updating the seed pool of $F_2$ by incorporating the global seed pool (i.e., the union set of seeds from all fuzzers).
\sys then runs $F_2$ for a certain cycle, in which each cycle performs the seed selection, energy assignment, and mutations of $F_2$ (\ding{194}).
After fuzzing with $F_2$, \sys updates the global seed pool with newly discovered seeds (\ding{195}) and recomputes the reward of individual fuzzers using our proposed fuzzer evaluation metric (\ding{196}). 
Finally, \sys updates the weight distribution based on the reward using our customized bandit model (Thompson sampling) (\ding{197}).
\sys continuously performs this iterative process until a predefined stopping criterion is reached.

Note that our design also addresses the first two limitations highlighted in \autoref{subsec:bg_rw}.
First, \sys distributes the given resources among the integrated fuzzers without needing extra resources.
Second, as detailed in Section~\ref{sec:impl}, \sys incorporates a global seed pool to enable real-time coverage sharing.
\begin{algorithm}[t]
    \caption{Assign reward to a selected fuzzer}
    \label{alg:evaluate}
    \SetAlgoLined
    \KwIn{The current fuzzing round $t$, the selected fuzzer $F_t$, the set of new seeds generated by this fuzzer $\mathcal{S}_t$, the maximum and minimum reward $r_{\text{max}}$ and $r_\text{min}$, and the set $\mathcal{I}$, recording the round at which each basic block is \textit{firstly} discovered.}
    $r_t \leftarrow 0$\;
    \For{seed $S$ in $\mathcal{S}_t$}{
        \For{new branch $b$ covered by $S$}{
        Suppose the branch $b$ goes from the basic block $BB_p$ to $BB_e$\;
        Find the round that $BB_p$ is firstly discovered in $\mathcal{I}$, noted as $t_p$;
        Compute the coverage interval $c$ $\leftarrow$ $t - t_p $\;
        Update the reward $r_t \leftarrow r_t + c$\;
        \If{$BB_e$ is newly discovered}{
        Add the pair $BB_e$ and $t$ to $\mathcal{I}$;
        }
            }
        }
    $r_{\text{max}} \leftarrow max(r_{\text{max}}, r_t)$\;
    $r_{\text{min}} \leftarrow min(r_{\text{min}}, r_t)$\;
    $r_t \leftarrow (r - r_{\text{min}}) / (r_{\text{max}} - r_{\text{min}})$\;
    \KwOut{normalized reward $r_t$}
\end{algorithm}

\subsection{Fuzzer Evaluation Methodology}
\label{subsec:tech_measure}

As described above, we introduce a novel fuzzer evaluation approach designed to gauge the performance of fuzzers and subsequently allocate rewards accordingly. In our framework, a high reward signifies a fuzzer's effectiveness. Below, we detail our evaluation methodology and the associated reward scheme.

To evaluate a fuzzer's performance, we consider not only its capability to uncover new branches but also the effort required to achieve such discoveries. This consideration stems from an observation that, during the initial stages of fuzzing, individual fuzzers generate a large number of seeds, contributing to an increase in coverage. However, as time progresses, identifying new branches becomes increasingly difficult, often leading to a stagnation in coverage growth. At this point, while most fuzzers face challenges in discovering new branches despite numerous mutation attempts, a select few manage to overcome complex constraints, generating seeds that unlock critical branches (as shown in \autoref{fig:coverage_example}). This breakthrough leads to a rapid emergence of new seeds, facilitated by the resolution of complex constraints and the subsequent ease of discovering additional branches.

Following this observation, we design our fuzzer assessment mechanism and reward scheme as follows. After running the selected fuzzer $F_t$ at the fuzzing round $t$, we obtain a set of new seeds, represented as $\mathcal{S}_t$.
The reward of the selected fuzzer, denoted as $r_t$, is equal to the accumulated rewards of all seeds $S \in \mathcal{S}_t$.
Each seed's reward is subsequently calculated from the accumulated rewards of all unique and newly covered branches ($b$) that are associated with it.
As indicated from Line 4 to Line 6 in~\autoref{alg:evaluate}, each branch's reward corresponds to its coverage interval ($c$). The latter quantifies how many rounds were required to discover this new branch.
Mathematically speaking, coverage interval $c$ equals the difference between the current round number ($t$) and when its preceding block (denoted as $BB_p$) was \textit{firstly} discovered.
We calculate $r_t$ by summing up values of $c$ over all newly covered branches across all seeds in $\mathcal{S}_t$.
Finally, we normalize $r_t$, using min-max scaling, to determine the final reward for fuzzer $F_t$ (Line 12-14). 
This approach allows us not only to capture overall coverage through the accumulation of newly discovered branches but also to account for the effort required via the computation of individual branches' coverage interval $c$.


\begin{figure}[t]
    \centering
    \includegraphics[width=0.8\linewidth]{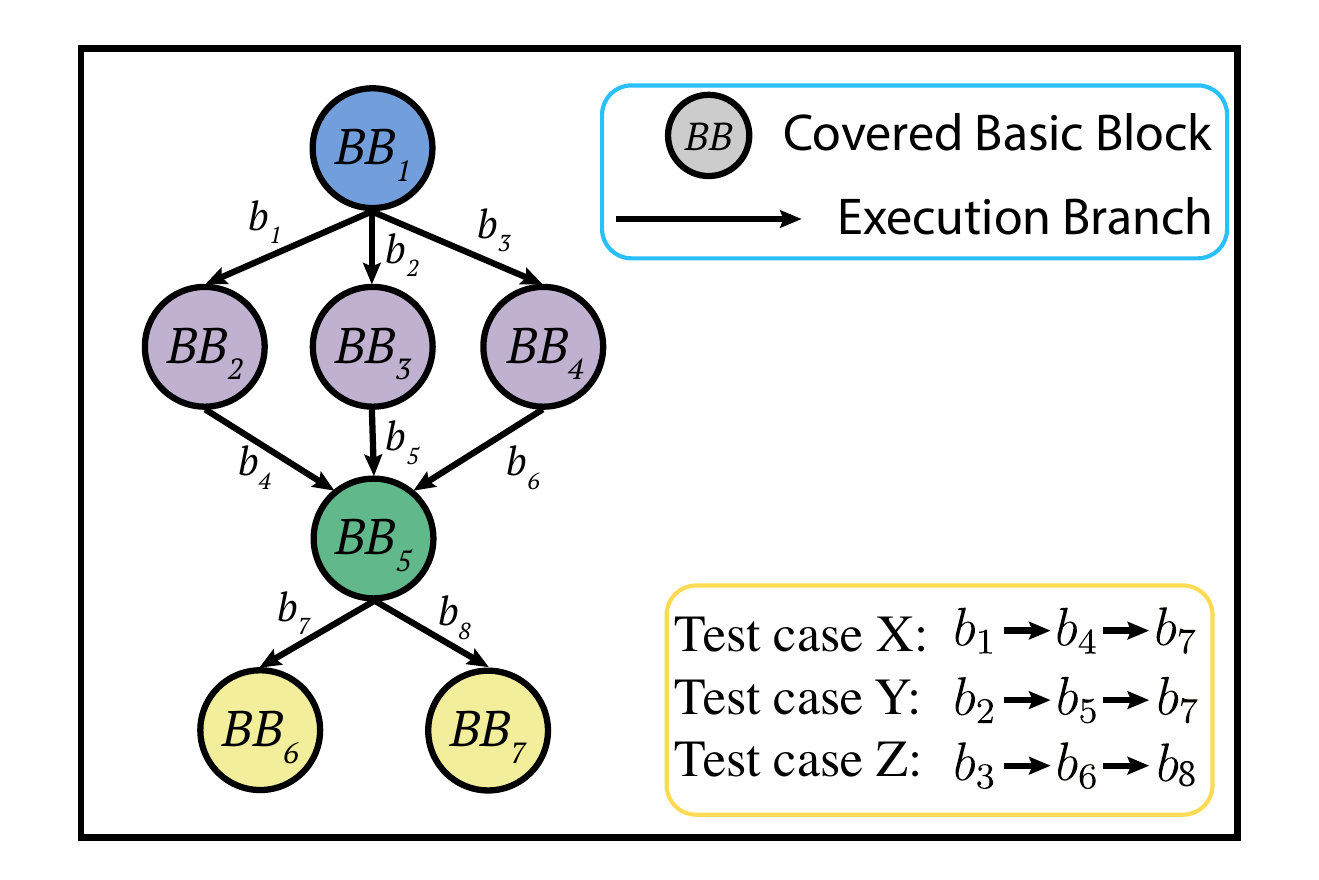}
    \caption{Motivating example for demonstrating reward computing.}
    \label{fig:evaluate}
    \vspace{-2mm}
\end{figure}

\noindent{\bf Example of reward computing.}
\label{appendix:reward_example}
As demonstrated in~\autoref{fig:evaluate}, the program features a target program with eight distinct branches.
Initially, seed $X$ identifies branches $b_1$, $b_4$, and $b_7$ during iteration $I_X$, leading to the inclusion of $\{BB_1=I_X, BB_2=I_X, BB_5=I_X, BB_6=I_X\}$ in set $I$.
Following a series of iterations without new discoveries, seed $Y$ emerges in round $t=I_Y$, detecting branches $b_2$, $b_5$, and $b_7$, with $b_2$ and $b_5$ being newly identified.
For branch $b_2$, $BB_1$ is recognized as the preceding block ($BB_p$) and $BB_3$ as the subsequent block ($BB_e$), resulting in a coverage interval calculation of $c = t - t_p = I_Y - I_X$. Consequently, $\{BB_3=I_Y\}$ is appended to set $I$.
Regarding $b_5$, $BB_3$ is the preceding block and $BB_5$ is the subsequent block, leading to a coverage interval of $c = I_Y - I_Y = 0$.
Since $BB_5$ was previously discovered, set $I$ remains unchanged.
The fuzzer that generates seed $Y$ is then awarded a reward $r_t = (I_Y - I_X) + (0) = I_Y - I_X$.
After additional rounds without findings, seed $Z$ is identified at round $t=I_Z$, uncovering new branches $b_3$, $b_6$, and $b_8$.
The coverage interval for $b_3$ is calculated as $c = I_Z - I_X$, with $\{BB_4=I_Z\}$ added to set $I$.
The interval for $b_6$ is $c = I_Z - I_Z = 0$, and for $b_8$, it is $c = I_Z - I_X$, leading to the inclusion of $\{BB_7 = I_Z\}$ in set $I$.
Ultimately, the fuzzer that generates seed $Z$ is awarded a reward $r_t = (I_Z - I_Y) + (0) + (I_Z - I_X) = 2I_Z - I_Y - I_X$.

\begin{algorithm}[t]
    \caption{\sys resource allocation}
    \label{alg:fuzzer_selection}
    \SetAlgoLined
    \KwIn{A set of individual fuzzer $\mathcal{F}$, global seed pool $\mathcal{S}_{g}$, reset interval $I_R$, time budget for each round $T_I$, stopping condition}
     $timer \leftarrow 0$, Number of round $t \leftarrow 0$\;
     \For{fuzzer $F$ in $\mathcal{F}$}{
     $F.total\_duration \leftarrow 0$, $F.num\_selection \leftarrow 0$ \;
     $F.cycles \leftarrow 1$\;
     $\alpha_F$ $\leftarrow$ 1, $\beta_F$ $\leftarrow$ 1\;
     }
     \While{stopping condition is not met}{
        Current fuzzer $ F_t \leftarrow$ TS($\mathcal{F}$) (\ding{192})\;
        Synchronize seeds from $\mathcal{S}_g$ to the seed pool of $F_t$ (\ding{193})\;
        $start \leftarrow$ current time\;
        \For{$j$=0, $j$<$F_t.cycles$, $j$++}{
            $F_t$.run\_one\_cycle() (\ding{194})\;
        }
        Update $\mathcal{S}_{g}$ with $\mathcal{S}_t $, new seeds generated by $F_t$  (\ding{195})\; 
        $d_t \leftarrow$ current time $- start$, $r_t \leftarrow$ evaluate $F_t$ with~\autoref{alg:evaluate}  (\ding{196})\;
        \tcp{Update probability distribution (\ding{197})} 
        $\alpha_{Ft}  \leftarrow \alpha_{Ft} + \hat{r}_t$, $\beta_{Ft}  \leftarrow \beta_{Ft} + (1 - \hat{r}_t)$\;
        \tcp{Auto-cycle}
        $F_t.total\_duration \leftarrow F_t.total\_duration+d_t$\;
        $F_t.num\_selection \leftarrow F_t.num\_selection + 1$ \;
        $F_t.avg\_cycle\_time$ $\leftarrow$ $ \text{floor} (\frac{F_t.total\_duration}{F_t.num\_selection}$)\;
        $F_t.cycles \leftarrow \frac{T_I}{F_t.avg\_cycle\_time}$\;
        \tcp{Reset mechanism}
        $timer \leftarrow timer + d_t$, $t \leftarrow t+1$ \;
        \If{$timer \geq I_R$}{
            $timer \leftarrow 0$,               $\alpha_F$ $\leftarrow$ 1, $\beta_F$ $\leftarrow$ 1 for every fuzzer\;
            }
        }
\end{algorithm}

\subsection{Resource Allocation Mechanism}
\label{subsec:tech_allocate}

Our resource allocation algorithm is shown in \autoref{alg:fuzzer_selection}.
After initializing the required variables (Lines 1-6), in each fuzzing round, our first step is to select a fuzzer $F_t$ using our customized bandits model (Line 8), followed by synchronizing its seeds with those in the global seed pool (Line 9).
Subsequently, we execute $F_t$, gather its reward, and update both its probability distribution and the global seed pool (Lines 10-16).
We also employ the auto-cycle mechanism to revise how many cycles a fuzzer should run when it is selected next time (Lines 17-20).
Finally, after updating the model for certain rounds, we reset parameters within our bandits model to ensure optimal performance moving forward (Lines 21-24).
In the following, we introduce our customized bandits model and the auto-cycle method.

\noindent{\bf Customized bandits model.}
Each arm represents an individual fuzzer $F\in \mathcal{F}$, associated with a weight distribution. 
Our goal is to keep updating the weight distributions in a way that maximizes the long-term accumulated reward obtained through repeated pulls of the arms.
By maximizing the long-term reward, the bandits can select the fuzzer benefiting the overall fuzzing campaign.
In contrast, the greedy method selects the fuzzers with the highest reward at each time but may not result in an optimal long-term reward.

We use the Thompson sampling algorithm (TS)~\cite{thompson1933likelihood} to update the weight distributions.
TS has both the SOTA empirical performance and theoretical guarantee in various problems~\cite{russo2018tutorial,chapelle2011empirical}. 
Suppose the fuzzer set $\mathcal{F}$ contains $K$ fuzzers.
For each fuzzer (or arm) $F_k$, TS defines the conditional distribution of the reward as a Bernoulli distribution parameterized by $\theta_{Fk}$.
Recall that in \autoref{subsec:tech_measure} we use $r_t$ to denote the reward of $F_t$. 
So $r_t | F_k \sim \text{Bern}(\theta_{Fk})$.
TS then defines the distribution of $\theta_{Fk}$ as a Beta distribution, parameterized by $\alpha_{Fk}$ and $\beta_{Fk}$, i.e., $\theta_{Fk} \sim \text{Beta}(\alpha_{Fk}, \beta_{Fk})$.
This is the weight distribution for $F_k$.
In each fuzzing round $t$, TS first samples $\theta_{Fk}$ from the current weight distribution for each fuzzer.
Then, it selects the fuzzer with the largest $\theta_{Fk}$, i.e., $F_t = \text{argmax}_{Fk} \theta_{Fk}$.
After running the selected fuzzer and collecting the reward $\hat{r}_t$, TS then updates the parameters of the weight distribution of the selected fuzzer as follows:
\begin{equation}
\alpha_{Ft} \leftarrow \alpha_{Ft} + \hat{r}_t, \ \ \beta_{Fk} \leftarrow \beta_{Fk} + 1 - \hat{r}_t \, .
\label{eq:beta}
\end{equation}

\begin{algorithm}[t]
    \caption{Customized TS bandit algorithm}
    \label{alg:customized-ts}
    \SetAlgoLined
    \KwIn{}
    $r_t \leftarrow 0$\;
    \For{\(t = 1, 2, \ldots\)}{
      \tcp{sample model}
      \For{\(k = 1\) \KwTo \(K\)}{
        Sample \(\hat{\theta}_k \sim \text{Beta}(\alpha_k, \beta_k)\)
      }
      \tcp{select and apply action}
      \(x_t \leftarrow \arg\max_k \hat{\theta}_k\) \\
      Apply \(x_t\) and observe \(r_t\) \\
      Sample \(\hat{r_t} \sim \text{Bern}(r_t)\) \\
      \tcp{update distribution}
      \((\alpha_{x_t}, \beta_{x_t}) \leftarrow (\alpha_{x_t} + \hat{r_t}, \beta_{x_t} + 1 - \hat{r_t})\)
    }
\end{algorithm}
%

Directly applying the original TS introduces two challenges. 
First, TS assumes $\hat{r}_t\in\{0, 1\}$ is discrete, while our reward falls in a continuous range of $[0, 1]$. 
To overcome this discrepancy, we employ a method in~\cite{agrawal2012analysis} to discretize our continuous reward $r_t$ into a binary equivalent $\hat{r}_t$. 
This discretized reward is then used to update the parameters following \autoref{eq:beta}, aligning our implementation with the TS framework.

TS also assumes that the reward distributions remain stationary over time.
However, in our scenario, as more program branches are discovered over time, these reward distributions will change accordingly. 
To account for such variations, we periodically reset each fuzzer's weight distribution parameters. 
This approach provides flexibility in capturing changes in the evolving reward distributions (see ~\autoref{appendix:thompson_example} for more details).

\noindent{\bf Auto-cycle mechanism.}
After choosing a fuzzer in each round, it is also necessary to determine the time and resources allocated to it. 
The guiding principle is fairness; we aim to assign similar resources in each round. 
It aligns with the bandits' assumption that rewards are obtained under similar resource conditions in every round.
A straightforward approach is to allocate a fixed run-time for each round. 
Unfortunately, this will break the fuzzing procedure of individual fuzzers. 
Instead, we specify the number of fuzzing cycles to run (one cycle means the full procedure from seed selection to generating new seeds), which better maintains the fuzzing procedure. 
Assigning the same number of cycles is also not an ideal approach, as some fuzzers operate faster than others, leading to unfairness.
To address this issue, we introduce an auto-cycle mechanism that dynamically adjusts the number of fuzzing cycles for each fuzzer based on its previous execution time.
Specifically, for each fuzzer, we first calculate its average cycle time ($F.avg\_cycle\_time$). 
We then establish a fixed time budget $T_I$, which determines how long a fuzzer can run in any given round. 
The number of cycles is computed as $\frac{T_I}{F.avg\_cycle\_time}$.

\section{Implementation}
\label{sec:impl}

In this section, we provide a comprehensive overview of the implementation, focusing on the integration, management, synchronization, and evaluation of fuzzers.

\subsection{Fuzzer Integration}
\label{subsec:impl_integration}

\sys currently supports 10 individual fuzzers, namely:
AFL~\cite{zalewski2020afl}, 
AFL++~\cite{fioraldi2020afl++}.
AFLFast~\cite{bohme2016CoveragebasedGreyboxFuzzing}, 
Angora~\cite{chen2018angora}, 
Darwin~\cite{jauernig2023DARWINSurvivalFittest}, 
FairFuzz~\cite{lemieux2018FairFuzzTargetedMutation}, 
honggfuzz~\cite{google2020honggfuzz}, 
LAF-Intel~\cite{Lafintel2016}, 
MOPT~\cite{lyu2022MOPTOptimizedMutation} and
Radamsa~\cite{2023AkiHelinRadamsa}.
These fuzzers were selected because of their status as state-of-the-art tools that incorporate diverse strategies to enhance various components of fuzzing. 
These aforementioned fuzzers operate on four different frameworks.
Specifically, the AFL~\cite{zalewski2020afl} framework underpins the operation of AFL, AFLFast, Darwin, and FairFuzz.
Meanwhile, AFL++, LAF-Intel, MOPT, Radamsa are based on the AFL++ framework~\cite{fioraldi2020afl++}.
Angora and honggfuzz each utilize their own unique frameworks.
Note that we enable \texttt{cmplog} within AFL++, a technique analogous to that utilized in RedQueen. 
\texttt{cmplog} is integrated as the default mode for AFL++.

In order to integrate these disparate fuzzers into a unified runtime environment, necessary but minimal modifications have been made to their underlying frameworks.
For example, fewer than 200 lines of code in both the AFL and AFL++ frameworks have been modified.
It is important to note that these integrated fuzzing frameworks are among the most widely used in this field. They serve as the platforms for many fuzzers ~\cite {yue2020ecofuzz, bohme2017directed, yun2018qsym, kim2020maxafl, li2019cerebro, lyu2022slime,zhang2022mobfuzz,zheng2023fishfuzz}.
This widespread usage suggests that \sys can effortlessly integrate additional fuzzers with existing modifications.

\subsection{Fuzzer Management, Synchronization \& Evaluation}
\label{subsec:impl_synchronization}
\sys uses a central communication interface that functions as a bridge to connect individual fuzzers. This interface is responsible for implementing most functionalities of \sys and coordinating with each fuzzer.

\noindent{\bf Fuzzer management.}
The central communication interface governs the operations of each fuzzer, including initiation, pausing, and resumption of their fuzzing tasks. It also monitors and reports on the runtime status of every fuzzer (e.g., determining when a specific fuzzer is engaged in seed selection). 
During their operational cycles, individual fuzzers may encounter two potential issues. The first issue arises when they become trapped in an infinite loop, resulting in an unexpectedly prolonged execution time for their current task. The second problem occurs if they fail and shut down during operation.
These complications can suspend \sys and negatively impact its overall performance. If the central interface detects that a currently active fuzzer has taken an unusually long time without completing its assigned cycles, it will dispatch a skip instruction to that particular fuzzer. This command pauses the fuzzer's operation and returns the current results to the central interface.
In instances where a fuzzer fails prematurely before completing its designated cycles, the central interface will send out a restart instruction to reboot it. Consequently, this allows for continuation toward the completion of its assigned fuzzing cycles.
These dual mechanisms ensure all fuzzers adhere strictly to round-time constraints $T_I$, even under circumstances involving infinite loops or failures.

\noindent{\bf Real-time synchronization.}
Simultaneously, this interface receives discoveries (i.e., seeds) from all connected fuzzers.
As previously detailed in~\autoref{subsec:tech_overview}, \sys operates by maintaining a global seed pool. The first step when selecting a fuzzer involves synchronizing its local seed pool with the global one. This synchronization process is achieved by tracking the local seed files present in the file system and identifying any missing seeds.
Given that the central interface has knowledge about each individual fuzzer's status, it can efficiently transmit these missing seeds to the chosen fuzzer prior to any form of seed selection.
Furthermore, we use this same method to monitor changes in the file system in order to enrich our global seed pool with new seeds. Consequently, newly discovered seeds are transferred from selected fuzzers into our global pool via this central interface.

\noindent{\bf Real-time evaluation.}
Once the central interface acknowledges the new seeds produced by the selected fuzzer, \sys incorporates them into set $\mathcal{I}$, as delineated in~\autoref{alg:evaluate}. 
Subsequently, it computes the reward for the selected fuzzer through an evaluation of these newly produced seeds. This computation is facilitated using a specially customized instrumentation binary equipped with coverage sanitizer and a customized runtime shared object that executes ~\autoref{alg:evaluate}.
\section{Experiment}
\label{sec:eval}

In this section, we comprehensively evaluate \sys to answer the following research questions (RQs):

\noindent{\bf RQ1.} \sys integrates a suite of advanced fuzzing tools, leveraging their combined strengths. 
When provided with identical computational resources, how does the collaborative fuzzing approach of \sys compare against the performance of the individual fuzzers it incorporates?

\noindent{\bf RQ2.} When compared to the state-of-the-art collaborative fuzzing method, \autofz, does \sys, enhanced with our proposed fuzzer evaluation and resource allocation method, demonstrate better performance?

\noindent{\bf RQ3.} \sys introduces three key techniques, including real-time synchronization via a global seed pool, fuzzer evaluation, and resource allocation. 
How do these key components individually contribute to the overall effectiveness of \sys?

\noindent{\bf RQ4.} To what extent does our customized bandits model improve the overall performance of \sys?

\noindent{\bf RQ5.} Whether \sys is sensitive to the key hyper-parameters (time budget for each round $T_I$ and the
reset interval $I_R$)?

In the following, we specify our experiment setup, as well as the design and results of the experiment to answer each research question. 

\subsection{Experiment Setup}
\label{subsec:exp_setup}

\noindent{\bf Experiment environment.}
We conduct all experiments on two workstations, each equipped with an AMD EPYC 7763 64-core processor and 512GB of RAM. 
Within this setup, each fuzzer operates in a separate Docker container (note that \sys is considered a single fuzzer).
Each docker container (fuzzer) is allocated with a single CPU core and 2GB of shared memory, ensuring a standardized, controlled, and fair testing environment.

\noindent{\bf Configuration of \sys.}
Recall that \sys periodically reset the bandits parameters to adapt to changes in reward distributions. 
Also, \sys requires specifying an appropriate time budget for the fuzzer selected by the bandits during the auto-cycle. 
The default setup for these two hyperparameters is that the reset interval $I_R$ is 120 mins, and the time budget $T_I$ is 120 seconds for each round.

\noindent{\bf Benchmark and fuzzing targets.}
We select the fuzzing targets in the benchmark database FuzzBench~\cite{metzman2021fuzzbench}.
We also use the Fuzzer Test Suite (FTS)~\cite{fuzzer_test_suite_2021} as an alternative benchmark because some fuzzers are not compatible with the targets in FuzzBench. 
Each fuzzer, representing both our proposed solution and established fuzzing techniques, was subjected to a continuous testing period of 24 hours for every fuzzing target. 
To mitigate inherent variability and ensure robust results, we executed each experiment~\textit{10 times}, resulting in a cumulative computation time of 7.5 CPU years.

\noindent{\bf Metrics.}
We select the widely used branch coverage as the our metric for evaluating the effectiveness of a fuzzer against a specific target program. 
Considering the varying performance that different fuzzers show across different programs, we calculate the average score for each fuzzer across different programs to enable a more comprehensive comparison.
The average score is introduced by FuzzBench~\cite{metzman2021fuzzbench}, which serves as a standard metric to evaluate the overall performance of a fuzzer. This metric allows for a more balanced comparison by accounting for the performance difference across multiple target programs, thereby offering a comprehensive view of a fuzzer's effectiveness. 
More specifically, the fuzzer's score for each target equals dividing its median branch coverage by the maximum branch coverage recorded for that same target.
Note that the highest relative performance does not necessarily correspond to 100\%.

\subsection{Experiment Design}
\label{subsec:exp_design}

\noindent{\bf Experiment I: Comparison with individual fuzzers.}
To address RQ1, we isolate the fuzzers incorporated within \sys, treating them as individual fuzzers.
We then run \sys and each individual fuzzer against the targets from FuzzBench.
As mentioned in~\autoref{subsec:exp_setup}, we run each fuzzer 24 hours with the same identical computational resources. 
We compute the branch coverage for each method in each individual target and the mean coverage across all targets.
We run each method 10 times and report the mean. 
We expect that \sys will have higher overall fuzzing performance compared to individual fuzzers, as \sys's can dynamically evaluate and select the most efficient fuzzers for different targets over time.

\noindent{\bf Experiment II: Comparison with \autofz.}
To address RQ2, we compare \sys with the state-of-the-art collaborative fuzzing framework \autofz.
It integrates 11 distinct fuzzers.
In Experiment I, we use the fuzzing targets in FuzzBench.
However, the implementation of \autofz is incompatible with FuzzBench.
As such, we use FTS as the benchmark for this experiment. 
Note that despite both \autofz and \sys incorporate multiple fuzzers, the composition of these fuzzers is not identical, featuring unique fuzzers such as Darwin in \sys and libFuzzer in \autofz.
To enable an apple-to-apple comparison, we restrict the individual fuzzers to those mutually supported by both \autofz and \sys. 
Specifically, we craft \autofz-8 and \sys-8 that integrate the following eight fuzzers: AFL, AFL++, AFLFast, LAF-Intel, MOpt, Radamsa, Angora, Fairfuzz. 
We compare the branch coverage between \autofz-8 and \sys-8 to evaluate which method has the better collaboration strategy on the same set of fuzzers. 

We also compared \sys-8 with the complete setup of \autofz, which includes all 11 fuzzers. 
This comparison aimed to evaluate the strengths of \sys against a well-established benchmark in the field. 
Demonstrating that \sys-8 outperforms the full \autofz would further support the effectiveness of \sys. 

\begin{table*}[t]
\centering
\resizebox{1\textwidth}{!}{
\begin{tabular}{lrrrrrrrrrr}
\toprule
Benchmark & \sys & AFL++ & honggfuzz & LAF-Intel & Radamsa & MOpt & AFL & AFLFast & FairFuzz  & DARWIN \\
\midrule
bloaty\_fuzz\_target & \cellcolor{lightgreen} 97.97 & \cellcolor{lightgreen} 97.41 & 77.56 & 96.38 & 82.45 & 94.95 & 93.65 & 93.78 & 80.52 & \dag 68.02 \\
curl\_curl\_fuzzer\_http & \cellcolor{lightgreen} 99.07 & \cellcolor{lightblue} 98.09 & 85.09 & 85.23 & 89.74 & 92.03 & 90.64 & 89.46 & 83.97 & \dag 62.99 \\
freetype2\_ftfuzzer & \cellcolor{lightgreen} 98.15 & \cellcolor{lightblue} 88.38 & 53.91 & 77.56 & 53.46 & 61.38 & 59.50 & 58.66 & 58.42 & \dag 21.27 \\
harfbuzz\_hb-shape-fuzzer & \cellcolor{lightgreen} 98.87 & \cellcolor{lightgreen} 98.84 & 81.73 & 97.22 & 84.47 & 96.42 & 95.74 & 95.34 & 88.69 & \dag 68.31 \\
jsoncpp\_jsoncpp\_fuzzer & \cellcolor{lightgreen} 100.00 & \cellcolor{lightgreen} 100.00 & \cellcolor{lightgreen} 100.00 & \cellcolor{lightgreen} 100.00 & \cellcolor{lightgreen} 100.00 & \cellcolor{lightgreen} 100.00 & \cellcolor{lightgreen} 100.00 & \cellcolor{lightgreen} 100.00 & \cellcolor{lightgreen} 100.00 & \cellcolor{lightgreen} 100.00 \\
lcms\_cms\_transform\_fuzzer & \cellcolor{lightgreen} 97.95 & \cellcolor{lightblue} 93.03 & 30.30 & 64.93 & 33.42 & 29.39 & 29.01 & 27.32 & 64.19 & 30.61 \\
libjpeg-turbo\_libjpeg\_turbo\_fuzzer & \cellcolor{lightgreen} 82.74 & \cellcolor{lightgreen} 82.65 & \cellcolor{lightgreen} 82.29 & \cellcolor{lightgreen} 82.65 & \cellcolor{lightgreen} 82.61 & \cellcolor{lightgreen} 82.26 & \cellcolor{lightgreen} 82.18 & \cellcolor{lightgreen} 82.19 & \cellcolor{lightgreen} 82.19 & \dag 81.64 \\
libpcap\_fuzz\_both & \cellcolor{lightgreen} 92.32 & \cellcolor{lightblue} 89.85 & 65.84 & 83.64 & 68.78 & 1.08 & 1.08 & 1.08 & 1.08 & \dag 0.00 \\
libpng\_libpng\_read\_fuzzer & \cellcolor{lightgreen} 95.95 & \cellcolor{lightgreen} 95.50 & 94.62 & \cellcolor{lightgreen} 95.52 & 92.47 & 93.07 & 93.02 & 92.12 & 93.19 & 93.26 \\
libxml2\_xml & \cellcolor{lightgreen} 99.61 & \cellcolor{lightgreen} 99.41 & 94.74 & - & 92.00 & 96.73 & 96.64 & 96.77 & 89.74 & 96.99 \\
libxslt\_xpath & \cellcolor{lightgreen} 98.31 & \cellcolor{lightgreen} 98.11 & 93.23 & 49.67 & 90.48 & 89.58 & 90.80 & 90.37 & 92.29 & 92.63 \\
mbedtls\_fuzz\_dtlsclient & \cellcolor{lightgreen} 74.73 & \cellcolor{lightgreen} 74.25 & 71.52 & 50.41 & 64.34 & 71.03 & 70.70 & 68.52 & 73.73 & \dag 49.30 \\
openh264\_decoder\_fuzzer & \cellcolor{lightgreen} 99.03 & 98.43 & 96.51 & \cellcolor{lightgreen} 99.28 & 79.75 & \cellcolor{lightgreen} 99.07 & \cellcolor{lightgreen} 99.06 & \cellcolor{lightgreen} 99.40 & 98.46 & \dag 0.00 \\
openssl\_x509 & \cellcolor{lightgreen} 99.95 & \cellcolor{lightgreen} 99.94 & 98.97 & \cellcolor{lightgreen} 99.59 & \cellcolor{lightgreen} 99.29 & \cellcolor{lightgreen} 99.69 & \cellcolor{lightgreen} 99.73 & \cellcolor{lightgreen} 99.72 & \cellcolor{lightgreen} 99.69 & \cellcolor{lightgreen} 99.91 \\
openthread\_ot-ip6-send-fuzzer & \cellcolor{lightblue} 77.05 & \cellcolor{lightgreen} 89.91 & 70.08 & 71.28 & 69.06 & 73.10 & 72.04 & 70.95 & 69.14 & \dag 0.00 \\
proj4\_proj\_crs\_to\_crs\_fuzzer & \cellcolor{lightgreen} 96.70 & \cellcolor{lightblue} 90.67 & 50.13 & 67.82 & 52.22 & 9.83 & 10.32 & 9.35 & 10.47 & \dag 0.70 \\
re2\_fuzzer & \cellcolor{lightgreen} 99.84 & \cellcolor{lightgreen} 99.43 & 98.39 & \cellcolor{lightgreen} 99.76 & 98.54 & 98.89 & \cellcolor{lightgreen} 99.34 & 98.63 & \cellcolor{lightgreen} 99.13 & \cellcolor{lightgreen} 99.01 \\
stb\_stbi\_read\_fuzzer & \cellcolor{lightgreen} 96.18 & \cellcolor{lightblue} 95.59 & 92.34 & 90.28 & 89.01 & 89.56 & 90.35 & 89.81 & 89.85 & 90.62 \\
systemd\_fuzz-link-parser & \cellcolor{lightgreen} 98.76 & \cellcolor{lightgreen} 98.35 & 88.64 & 97.93 & 89.67 & 90.91 & 90.50 & 90.70 & 84.30 & \dag 51.65 \\
vorbis\_decode\_fuzzer & \cellcolor{lightblue} 98.79 & \cellcolor{lightblue} 98.40 & 97.62 & \cellcolor{lightgreen} 99.38 & 93.75 & 97.58 & 97.74 & 97.70 & 97.54 & \dag 59.25 \\
woff2\_convert\_woff2ttf\_fuzzer & \cellcolor{lightblue} 98.56 & 97.36 & 89.19 & \cellcolor{lightgreen} 99.26 & 82.26 & 90.10 & 88.86 & 89.69 & 81.56 & \dag 62.95 \\
\hline
Median Score & \cellcolor{lightgreen} 98.31 & \cellcolor{lightblue} 97.41 & 88.64 & 90.28 & 84.47 & 90.91 & 90.64 & 90.37 & 84.30 & 62.99 \\
Average Score & \cellcolor{lightgreen} 95.26 & \cellcolor{lightblue} 94.46 & 81.56 & 81.32 & 80.37 & 78.89 & 78.61 & 78.17 & 78.01 & 58.53 \\
\bottomrule
\end{tabular}
}
\caption{Fuzzer score summary in FuzzBench report. The fuzzers are organized based on average scores, representing their average relative coverage. Those with higher values are positioned toward the left side of the table.
A green background signifies a superior score, whereas a blue background denotes the second-best score. Note that in FuzzBench report, any difference less than 1\% is considered as a random effect and thus disregarded. The symbol \dag \ indicates that the fuzzer encounters a crash before completing its 24-hour task. The symbol - indicates that the fuzzer does not support the target.}
\label{tab:fuzzbench}
\end{table*}

\noindent{\bf Experiment III: Ablation study.}
We conduct a comprehensive ablation study to answer RQ3 (i.e., the efficacy of our proposed seed synchronization, fuzzer evaluation, and resource allocation mechanism). 
First, we configure \sys to operate without conducting seed synchronization with the global seed pool (denoted as ``no-sync'').
Comparing no-sync with \sys can quantify the importance of our seed synchronization mechanism. 
Our hypothesis is that without a synchronization mechanism, there is a possibility for the fuzzers in \sys to redundantly explore the same paths, leading to a waste of time and resources. 

Second, to evaluate the importance of our fuzzer evaluation method, we adjust the reward allocation for fuzzer $F$ in \autoref{alg:evaluate} by incrementing $\alpha_F$ by 1 upon any new coverage discovery and increasing $\beta_F$ by 1 when it fails. 
Additionally, we deactivated the auto-cycle feature to facilitate precise reward calculation. 
We denote this variation as ``naive-reward''.
Comparing this variation with the simple fuzzer evaluation method reveals the effectiveness of our proposed evaluation method that considers the difficulty of the resolved branches. 

Third, we substitute our bandits-based resource allocation mechanism with a simple random scheduling approach (denoted as ``random''). 
Comparing the performance of this variant against \sys enables us to quantify the impact of our bandits-based allocation mechanism on the efficacy of the fuzzing process.

We run \sys and the three variations on four targets from FuzzBench: \texttt{bloaty}, \texttt{freetype2}, \texttt{woff2ttf}, and \texttt{lcms}.
These targets were chosen due to their varied sizes and complexities, which provide a comprehensive array of challenges that a fuzzer may face when dealing with different code structures.
To elaborate, \texttt{bloaty} is a large-scale program with 89,530 edges and is equipped with 94 initial seeds. 
Conversely, \texttt{freetype2} has 19,056 edges but operates with only two seeds, representing a large program with a minimal initial seed count. 
\texttt{woff2ttf} is a smaller program with 10,923 edges and 62 initial seeds. 
Finally, \texttt{lcms} is the smallest program with 6,959 edges and just one seed, representing a compact program with a sparse seed allocation.

\noindent{\bf Experiment IV: Effectiveness of the customized bandits model.}
Recall that our customized bandits model introduces two key designs, reward discretization and periodic parameter reset.
Here, reward discretization cannot be changed because, without a discrete reward, we cannot use the TS algorithm for our method.
As such, to address RQ4, we remove the periodic parameter reset mechanism (denoted as ``no-reset'') and run in on the four targets used in Experiment III. 
Comparing this variation with \sys can demonstrate the efficacy of considering the dynamical changes in bandits' reward distribution. 

\noindent{\bf Experiment V: Hyper-parameter sensitivity.}
To answer RQ5, we vary two key hyper-parameters of \sys: the time budget for each round $T_I$ and the reset interval $I_R$.
Specifically, we adjust $T_I$ between 90 to 180 seconds in 30-second increments and vary $I_R$ from 90 to 180 minutes at 30-minute intervals.
We compare these variations with \sys's default setup on the four targets used in Experiment III \& IV to quantify the sensitivity of \sys against these hyper-parameters. 
We conduct a Nemenyi post-hoc test~\cite{hollander2013nonparametric}, which is widely used to determine whether there are notable result differences among different experimental setups.


\begin{table}[t]
\Large
\resizebox{0.47\textwidth}{!}{
\begin{tabular}{l c c c c c c }
\toprule \multirow{2}{*}{ Benchmarks } & \multicolumn{2}{c}{ \sys-8 } & \multicolumn{2}{c}{ \autofz-8 } & \multicolumn{2}{c}{ \autofz } \\
\cline{2-7}
 & mean & std & mean & std & mean & std \\
\midrule
boringssl-2016-02-12 & 1671.7 & 0.48 & 1622.9 & 12.03 & \textbf{1678.5} & 17.06 \\
c-ares-CVE-2016-5180 & 42.9 & 0.32 & \textbf{43.0} & 0.00 & 42.9 & 0.32 \\
freetype2-2017 & \textbf{10946.4} & 418.56 & 7887.1 & 326.99 & 8236.8 & 234.77 \\
guetzli-2017-3-30 & \textbf{1409.9} & 3.75 & 1392.9 & 9.39 & 1399.6 & 8.25 \\
harfbuzz-1.3.2 & \textbf{4866.0} & 41.63 & 4622.8 &57.72 & 4722.3 &53.50 \\
json-2017-02-12 & 649.1 & 1.52 & 649.5 & 5.82 & \textbf{662.9} & 0.32 \\
libarchive-2017-01-04 & \textbf{5071.1} & 190.69 & 4453.0 & 308.46 & 4687.7 & 300.56 \\
libjpeg-turbo-07-2017 & \textbf{2028.9} & 91.89 & 1836.7 & 79.63 & 2026.6 & 42.24 \\
libpng-1.2.56 & 1054.1 & 9.54 & 1076.9 & 11.86 & \textbf{1090.0} & 8.82 \\
libssh-2017-1272 & \textbf{1068.0} & 59.60 & 1006.9 & 14.50 & 1064.9 & 54.02 \\
libxml2-v2.9.2 & \textbf{8308.8} & 292.49 & 3955.4 & 73.62 & 4406.6 & 353.30 \\
lcms-2017-03-21 & 1923.8 & 179.52 & \textbf{1991.8} & 39.44 & 1976.1 & 37.25 \\
openssl-1.0.1f & \textbf{5536.4} & 551.51 & 5058.8 & 822.06 & 5450.7 & 250.61 \\
openssl-1.0.2d & \textbf{2136.1} & 1.66 & 2126.4 & 3.37 & 2133.1 & 2.60 \\
openssl-1.1.0c-bignum & \textbf{2036.9} & 0.74 & 2034.8 & 0.42 & 2034.7 & 0.82 \\
openssl-1.1.0c-x509 & 6105.8 & 0.63 & 6144.0 & 3.62 & \textbf{6145.3} & 1.70 \\
openthread-2018-02-27-ip6 & 1412.5 & 318.97 & 1280.4 & 323.10 & \textbf{1412.8} & 320.93\\
openthread-2018-02-27-radio & 2783.7 & 240.00 & 2595.2 & 290.29 & \textbf{3260.8} & 42.63 \\
pcre2-10.00 & \textbf{9909.4} & 125.82 & 8956.5 & 73.64 & 9124.1 & 118.82 \\
proj4-2017-08-14 & 2461.4 & 192.53 & 1015.1 & 595.85 & \textbf{3568.7} & 43.01 \\
re2-2014-12-09 & \textbf{2485.3} & 7.10 & 2470.6 & 8.85 & 2478.7 & 7.24 \\
sqlite-2016-11-14 & \textbf{1981.0} & 1112.83 & 1551.5 & 147.60 & 1624.0 & 0.00 \\
vorbis-2017-12-11 & \textbf{1264.1} & 6.94 & 1236.3 & 10.75 & 1242.4 & 12.66 \\
woff2-2016-05-06 & 1071.9 & 23.32 & \textbf{1097.1} & 16.93 & 1062.7 & 33.05 \\
wpantund-2018-02-27 & \textbf{3315.2} & 99.03 & 3079.3 & 84.01 & 3176.1 & 52.75 \\
\midrule 
Average Coverage Enhancement & \multicolumn{2}{c}{-} & \multicolumn{2}{c}{16.0\%} & \multicolumn{2}{c}{4.8\%} \\
Average Score & \multicolumn{2}{c}{ 91.5 } & \multicolumn{2}{c}{ 84.3 } & \multicolumn{2}{c}{  89.6 }\\
\bottomrule
\end{tabular}}
\caption{The branch coverage of \sys vs. autofz-8 and full autofz. Each row displays the arithmetic mean and standard deviation of branch coverage for every fuzzer. The maximum branch coverage achieved for each target is also highlighted.}
\label{tab:autofz}
\end{table}

\subsection{Experiment Result}
\label{subsec:exp_result}

\noindent{\bf Experiment I: Comparison with individual fuzzer.} 
\autoref{tab:fuzzbench} shows the average score of each fuzzer across individual target programs, as well as the median and mean score, after a 24-hour testing period. 
The table first shows that \sys achieves the highest median and average score across all the FuzzBench targets.
Regarding individual targets, \sys outperforms all other individual fuzzers in 18 of the 21 evaluated targets.
This result validates \sys's capability in harnessing and ensembling the unique strengths of individual fuzzers. 
In other words, \sys's bandit algorithm is able to consistently allocate more resources to fuzzers with the best performance along the fuzzing process. 
As a result, it puts together an ensemble strategy with the best performance. 

\autoref{fig:compare_individual} in the Appendix shows the branch coverage of \sys vs individual fuzzers across the time, we can observe that \sys is more stable than individual fuzzers.
This is another advantage of being an ensemble strategy in that \sys can always choose a functional fuzzer and will not be largely affected by the failure of one or two fuzzers. 
The fuzzer management feature introduced in \autoref{subsec:impl_synchronization} also helps the stability. 
This feature enables \sys to promptly identify and address any issues with the fuzzers during operation, ensuring consistent performance.

However, it is noteworthy that despite its overall superior performance, \sys occasionally exhibits a slower initial growth in code coverage during the initial 30 to 60 minutes compared to certain individual fuzzers (\autoref{fig:compare_individual} ). 
We believe this is an expected characteristic of \sys's operation, as \sys necessitates multiple iterations to effectively explore and determine the dynamic effectiveness of the various fuzzers it includes.

Finally, we notice that the coverage of score of \sys demonstrates relatively marginal improvement than AFL++. 
This is mainly because AFL++ (with \textit{cmplog} enabled) performs much better than other fuzzers in most targets. 
By analyzing the seeds, we found that AFL++ dominates other fuzzers in code coverage and is also assigned the most resources.
This demonstrates that \sys indeed identifies the best fuzzer and allocates most resources to it.  
Furthermore, we believe the similar performance of \sys and AF++  does not dilute the necessity of \sys for the following reasons.
First, as demonstrated in~\autoref{sec:real-world-assessment}, \sys performs much better than the state-of-the-art fuzzers in mutation scores, another widely used metric for evaluating a fuzzer's ability to find vulnerabilities. 
Second, as an ensemble approach, \sys is more stable than individual fuzzers like AFL++.
Third, in scenarios where AFL++ under-performs, \sys can still choose other superior fuzzers and achieve decent and stable performance.

\noindent{\bf Experiment II: Comparison with \autofz.}
\autoref{tab:autofz} presents a comparison between \sys-8, \autofz-8, and \autofz on FTS targets. 
First, the table shows that \sys-8 outperforms \autofz-8 in terms of average branch coverage, achieving an improvement of 16.0\%.
More specifically, \sys outperforms \autofz-8 in 19 of the 25 evaluated programs with a notable margin in branch coverage.
This superior performance validates the effectiveness of our advanced collaboration strategy, including our real-time seed synchronization, faithful and efficient fuzzer evaluation, and bandits-based resource allocation/fuzzer scheduling strategy.
In particular, it verifies our discussion in~\autoref{sec:bg} and~\autoref{sec:methods} that \autofz only realizes greedy resource allocation strategies while \sys can learn globally more optimal strategies. 

It is also worth noting that in cases where \sys-8 does not outperform \autofz-8, the decline in coverage is relatively marginal (not exceeding 3.5\%). 
The standard deviation of branch coverage for \sys-8 in these specific targets aligns closely with that of \autofz-8. 
Furthermore, it is important to highlight an outlier occurrence, where \sys-8's standard deviation on the target \texttt{sqlite-2016-11-14} was exceptionally high, because of an isolated trial that yielded remarkably high branch coverage.
This result shows that \sys has a higher stability than \autofz. 

Moreover, \sys also slightly outperforms the original \autofz, which incorporates three additional fuzzers (including some hybrid fuzzers). 
Despite having fewer fuzzers, \sys-8 achieves an average branch coverage that exceeds that of the original \autofz by 4.8\%. 
Although \sys does not integrate hybrid fuzzers (with concolic execution), powered with our advanced resource allocation method, \sys is superior to \autofz, which integrates these advanced fuzzers. 
This further demonstrates the importance of our bandits-based resource allocation method.

\begin{figure}[t]
    \centering
    \includegraphics[width=1\linewidth]{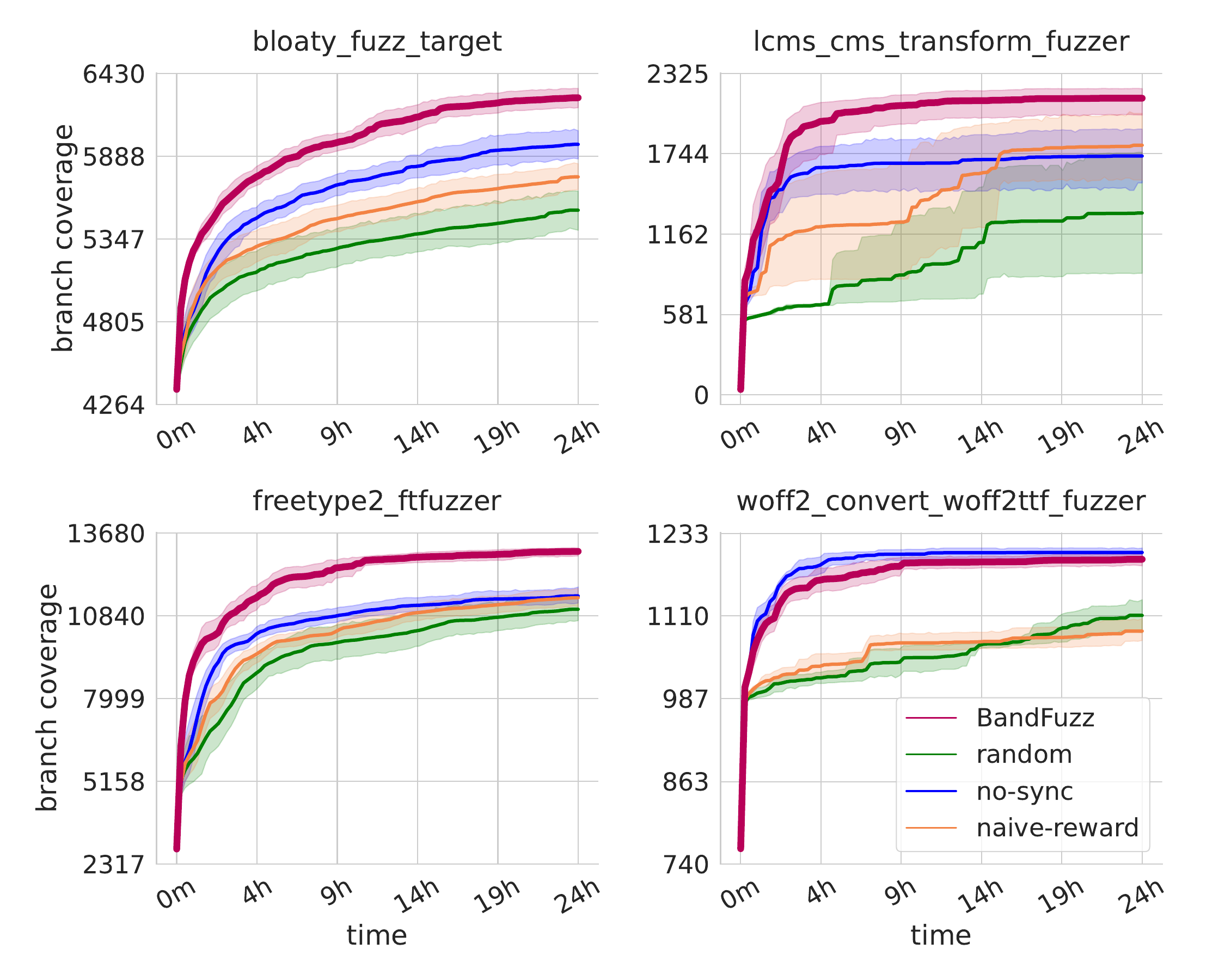}
    \caption{Comparison among \sys and two variants. Each line plot represents the arithmetic mean along with a 95\% confidence interval, derived from 10 independent trials. The label \textit{random} refers to \sys that employs random resource allocation in each round, and \textit{no-sync} represents \sys operating without seed synchronization.}
    \label{fig:ablation_collaborative}
    \vspace{-2mm}
\end{figure}

\noindent{\bf Experiment III: Ablation study.} 
\autoref{fig:ablation_collaborative} shows the ablation study results. 
First, \sys outperforms ``no-sync'' by an average of 9.6\% in branch coverage. 
This improvement validates the effectiveness of our real-time seed synchronization mechanism. 
It prevents individual fuzzers from repeatedly covering the same branches, enabling a more thorough and efficient exploration of the entire program space.

However, an exception is observed where \sys does not demonstrate superiority over ``no-sync'' in the case of \texttt{woff2}. 
This deviation is because of the unique characteristics of the target - a sparse edge count combined with a large number of initial seeds. 
In such cases, the advantage derived from inter-fuzzer knowledge exchange becomes less significant since each fuzzer can already have a comprehensive exploration of the program's structure.

Second, \autoref{fig:ablation_collaborative} also shows that \sys outperforms ``naive-reward'' by about 5.8\% in branch coverage. 
This result verifies the importance of our proposed fuzzer evaluation mechanism. 
As discussed in~\autoref{subsec:tech_overview}, assigning rewards solely based on coverage is limited in that it cannot account for the difficulty and complexity of each new coverage, resulting in a less comprehensive assessment of fuzzer performance. 

Last but not least, the ``random'' variation triggers the largest performance drop, recording about a 22.0\% reduction in branch coverage. 
This highlights the importance of our bandits-powered resource allocation strategy, which explore global optimal solutions for resource allocation, substantially optimizing fuzzing efficiency.

\begin{figure}[t]
    \centering
    \includegraphics[width=1\linewidth]{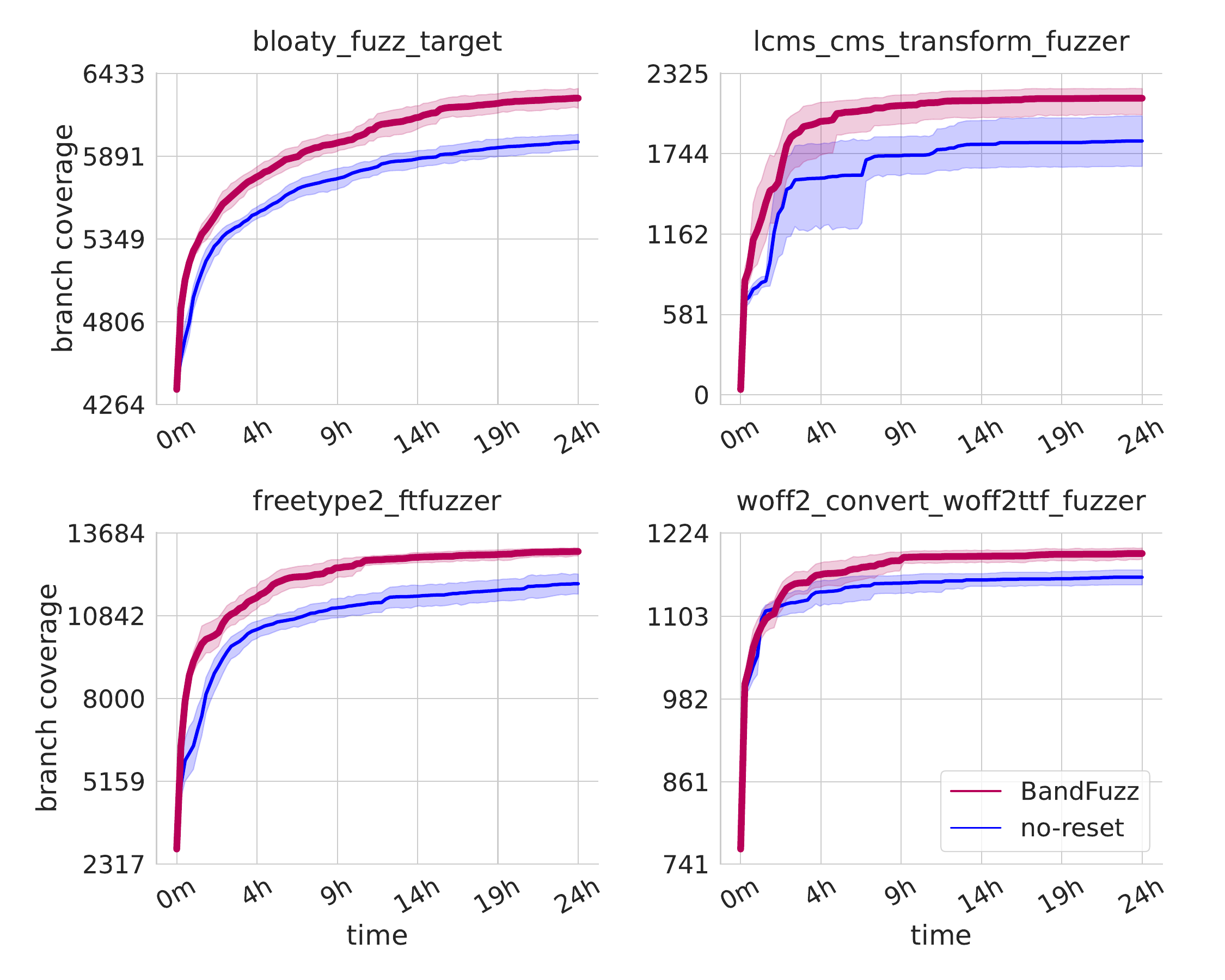}
    \caption{Comparison among \sys and two variants. Each line plot represents the arithmetic mean along with a 95\% confidence interval, derived from 10 independent trials. The label \textit{no-reset} refers to \sys without reset mechanism, and \textit{naive-reward} represents \sys adopting a naive algorithm to evaluate fuzzers.}
    \vspace{-2mm}
    \label{fig:ablation_ts}
\end{figure}

\begin{figure}[t]
    \centering
    \includegraphics[width=1\linewidth]{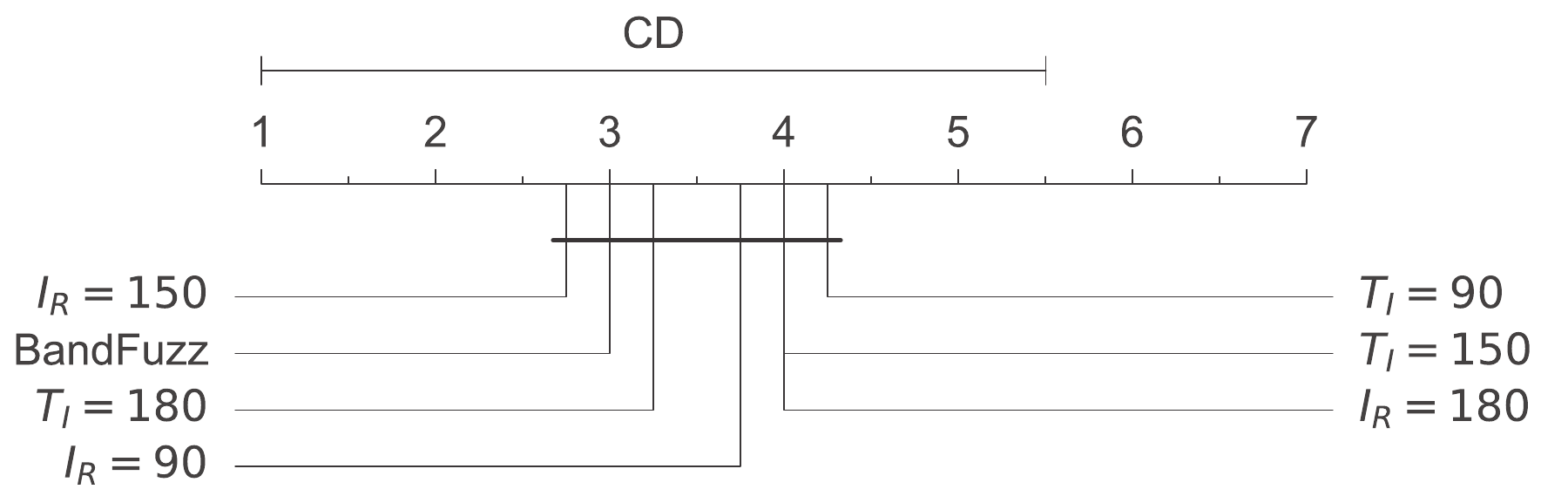}
    \vspace{-3mm}
    \caption{Critical Difference (CD) among various configurations of \sys. The average ranking of fuzzers is represented by each number. The horizontal line indicates no critical difference in performance among the grouped fuzzers, as determined by the Nemenyi post-hoc test. Each label represents a variation of the depicted hyperparameter while keeping another parameter at its default values.}
    \label{fig:ablation_san_cd}
    \vspace{-2mm}
\end{figure}

\noindent{\bf Experiment IV: Effectiveness of the customized bandits model.}
\autoref{fig:ablation_ts} demonstrates the comparison between \sys and the ``no-reset'' version of \sys, which excludes the parameter reset operation in our customized bandits model.
The figure shows an 8.1\% decrease in branch coverage for ``no-reset''. 
This result verifies the necessity of our customized bandits model.
As discussed in~\autoref{subsec:tech_allocate}, without a reset mechanism,  \sys' cannot mitigate the impact of outdated information.
As a result, ``no-reset'' cannot precisely adapt to changes in reward distributions and thus has limited fuzzing effectiveness. 

\noindent{\bf Experiment V: Hyper-parameter sensitivity.}
\autoref{fig:ablation_san_cd} depicts the performance differences introduced by varying two key hyper-parameters of \sys: $T_I$ and $I_R$.
The figure shows that changing these two hyper-parameters introduces only marginal changes in overall fuzzing effectiveness. 
The result verifies the insensitivity of \sys against key hyper-parameter changes.
This also demonstrates the practicality of \sys, as users are not required to carefully choose the optimal hyper-parameters.
\section{Real-World Assessment}
\label{sec:real-world-assessment}

Our evaluation of \sys extends beyond controlled experiments to its application in a recent, highly competitive fuzz testing event, SBFT 2024, as detailed in~\cite{anonymous-sbft-link}. 
This event presents a modified FuzzBench platform for evaluating advanced fuzzing techniques in a real-world context.

During the competition, participant fuzzers, including \sys, are deployed on the modified FuzzBench platform with the same amount of resources for the same period of time. 
Notably, the evaluation metric is different from our experiments in~\autoref{sec:eval}, which primarily utilize branch coverage as the performance metric. 
Instead, the competition organizers employ the~\textit{mutation analysis} as the primary evaluation tool, a method that is increasingly used as an alternative to traditional coverage metrics in fuzzing benchmarks, as discussed in~\cite{gorz2023systematic}.

Mutation analysis introduces synthetic faults into target programs to create program mutants, minimizing human bias in the process. 
A mutant is deemed ``killed'' if the fuzzer generates at least one input causing a deviation from the original program's behavior (i.e., trigger a synthetic fault).
The effectiveness of a fuzzer is thus quantified by the number of mutants it successfully killed, with a higher count indicative of superior performance.
As discussed in~\cite{gorz2023systematic}, this metric offers a more robust and comprehensive assessment compared to coverage by addressing issues like saturation and over-fitting. 
Moreover, by integrating a variety of fault types into the program, mutation analysis provides a broader spectrum evaluation of a fuzzer's capability to discover vulnerabilities, which is the main purpose of fuzzing in the security context.

The mutation analysis results, as presented in~\autoref{tab:killed}, clearly demonstrate \sys's outstanding performance in the competition. 
It surpassed not only widely recognized fuzzers like libAFL~\cite{fioraldi2022libafl} and libFuzzer~\cite{2018libfuzzer} but also advanced contenders such as FishFuzz~\cite{zheng2023fishfuzz} and Pastis~\cite{david2023pastis}, alongside two anonymously fuzzers denoted as \textit{AF1} and \textit{AF2}. 
Specifically, \sys achieved the highest count of killed mutants in six out of the eight targeted categories. 
Moreover, while the other fuzzers showed relatively narrow variations in their average mutation scores (ranging from 79 to 86), \sys recorded the highest average, a significant 98, marking a substantial lead over its competitors. 
The mutation score is calculated by taking the median number of mutants it killed and dividing it by the highest number of mutants killed recorded for the same target. 
In comparison to the second-ranked fuzzer, TuneFuzz~\cite{zheng2023fishfuzz}, \sys's average mutation score is higher by an impressive 14\%. 
Overall, this evaluation demonstrates the significant superiority of \sys over other SOTA fuzzers (widely used in academia or industry) in~\textit{bug detection capability for complex, real-world programs}.

\begin{table}[t]
\caption{ 
Number of mutants killed per benchmark, with each column reflecting the total mutants killed by the respective fuzzer.
}
\resizebox{0.48\textwidth}{!}{
\begin{tabular}{lccccccc}
\hline
Targets & \sys & FishFuzz & AF1 & Pastis & AF2 & libAFL & libFuzzer \\
\hline
freetype2\_ftfuzzer & \textbf{6814} & 5924 & 5944 & 6158 & 5900 & 6172 & 5131 \\
jsoncpp\_jsoncpp\_fuzzer & \textbf{1150} & 672 & 650 & 686 & 651 & 638 & 679 \\
lcms\_cms\_transform\_fuzzer & \textbf{1709} & 1461 & 1522 & 1560 & 1570 & 1547 & 1527 \\
libpcap\_fuzz\_both & 2095& 2096 & \textbf{2443} & 1852 & 2031 & 1811 & 1636 \\
libxml2\_xml & \textbf{7259} & 6582 & 6880 & 6240 & 6212 & 6160 & 6125 \\
re2\_fuzzer & \textbf{6651} & 5967 & 3672 & 3670 & 3650 & 3635 & 3687 \\
stb\_stbi\_read\_fuzzer & 1597 & 1643 & 1550 & \textbf{1647} & 1592 & 1562 & 1474 \\
zlib\_zlib\_uncompress\_fuzzer & \textbf{361} & 333 & 352 & 359 & 326 & 326 & 355 \\
\hline
Average Mutation Score & \textbf{98} & 86 & 84 & 82 & 81 & 79 & 77 \\
\hline
\end{tabular}}
\label{tab:killed}
\vspace{-2mm}
\end{table}

\section{Related Work}
\label{sec:rw}

In addition to heuristic-based and program analysis-based fuzzers mentioned in~\autoref{subsec:bg_fuzzer}, recent research also explores learning-based fuzzers~\cite{yue2020ecofuzz,wang2021reinforcement,lyu2022slime,koike2022slopt,yu2023gptfuzzer,deng2023large,xia2024fuzz4all,wang2021syzvegas,zong2020fuzzguard,cummins2018compiler,she2019neuzz,she2020mtfuzz,wu2022evaluating,nicolae2023revisiting,bottinger2018deep,liu2019deepfuzz,zhu2023better,chen2023rltrace,li2022alphaprog}. 
These methods leverage a wide range of ML methods, such as deep learning and reinforcement learning, to enhance specific components of existing fuzzers or to construct new fuzzing frameworks.
Among these methods, a notable line of research utilizes multi-arm bandits to improve existing fuzzers~\cite{yue2020ecofuzz, wang2021reinforcement, zhao2021evolutionary, lyu2022slime,zhang2022mobfuzz,wu2022one,scott2021banditfuzz}.
In what follows, we summarize these methods and discuss their limitations.

\noindent{\bf Improving seed selection.}
Existing bandits-based seed selection methods~\cite{yue2020ecofuzz, wang2021reinforcement, zhao2021evolutionary, lyu2022slime} typically treat the seeds as the arms and employ standard bandit algorithms, such as Upper Confidence Bound (UCB)~\cite{auer2002finite}, to dynamically adjust the weight of each seed based on the fuzzing feedback. 
While these techniques outperform heuristic-based strategies, they face two key limitations. 
First, dealing with a large volume of seeds results in bandits having a large number of arms, leading to a less efficient learning process.
Furthermore, the continual growth in the number of seeds violates the fundamental assumption of bandits, which is designed for a fixed number of arms.
Applying bandit algorithms without addressing these issues may lead to performance improvements that are not universally applicable across different test scenarios.

\noindent{\bf Enhancing power scheduling.}
Beyond seed selection, recent studies have also applied bandit algorithms to enhance power scheduling.
For example, Zhang et al.~\cite{zhang2022mobfuzz} treat seeds as the arms of bandits.
However, a notable difference is that researchers employ the UCB algorithm to dynamically optimize not just the selection of seeds but also the distribution of energy. 
Consequently, bandits-based power scheduling encounters similar challenges as those faced by bandits-based seed selection methods.

\noindent{\bf Optimizing input mutation.}
Recent research also leverages bandits for mutation scheduling~\cite{koike2022slopt,wu2022one,wang2021cmfuzz,karamcheti2018adaptive,scott2021banditfuzz}. 
For example, Wu et al.~\cite{wu2022one} designed a two-layer bandits. 
Here, the arms of the first layer are the number of mutators to be used and the second layer's arms are types of mutators.
This method leverages the UCB1-Tuned algorithm~\cite{auer2002finite} to select arms for both layers to guide mutator selection.
This method, although not a typical bandits design, demonstrates a decent empirical performance.  

In this work, rather than use it for improving individual components, we leverage multi-arm bandits to build a collaborative fuzzing framework, which is more robust and generic than individual fuzzers.

%
\section{Discussion}
\label{sec:discussion}


\noindent{\bf More fuzzer support.} 
\sys incorporates 10 state-of-the-art fuzzers, which employ advanced strategies to enhance nearly every fuzzing component.
Hybrid Fuzzing is the only main category of fuzzers that \sys has not employed.
These fuzzers incorporate concolic (\textbf{conc}rete symb\textbf{olic}) executors to improve the fuzzers' ability to solve difficult constraints.
Integrating concolic execution into \sys presents several challenges. 

First, synchronizing seeds between concolic executors and other fuzzers cannot be simply achieved via copying the global seed pool, as concolic executors are not aware of the coverage of fuzzers.  
Without synchronization, concolic executors will explore program states already founded by other fuzzers, leading to significant resource wastage. 
Additionally, given that concolic executors are much slower than regular fuzzers, it is difficult to maintain a fair resource allocation between fuzzers and concolic executors while capturing fuzzer dynamics in real-time. 

Tackling these challenges requires a comprehensive update of \sys, including designing a new seed synchronization and fuzzer evaluation mechanism, which demands non-trivial efforts. 
Our primary goal is to investigate an effective collaborative fuzzing strategy.
Additionally, we position \sys as the preliminary effort toward comprehensive collaborative fuzzing frameworks.
As such, we defer to integrating hybrid fuzzers as our future work.

\noindent{\bf Alternative seed assessment.}
Recall that \sys employs code coverage and the interval between branch discoveries as metrics to assess the quality of new seeds. 
However, there exist additional metrics that could be used to evaluate seed quality, such as basic-block centrality~\cite{she2022effective} and the distance to sanitizer~\cite{zheng2023fishfuzz}. 
In future work, we plan to explore other seed quality assessment metrics and refine our fuzzer evaluation accordingly.

\noindent{\bf TS improvement.}
Recall that we developed a customized Thompson Sampling (TS) approach for our bandits model, incorporating reward discretization to manage continuous rewards and a reset mechanism to tackle non-stationary reward distributions.
Although these modifications are effective, there is room for improvement or replacement with alternative techniques.
For instance, we could evaluate different discretization strategies or reformulate TS by treating rewards as a continuous distribution rather than a discrete Bernoulli distribution. 
Furthermore, exploring adversarial bandits, such as EXP3~\cite{auer2002nonstochastic}, which are specifically designed to handle non-stationary reward distributions, could offer additional enhancements to our model.

\noindent{\bf Efficient compilation.}
As a collaborative fuzzing framework, \sys is designed to integrate with any type of fuzzer. 
Technically, the fuzzers incorporated in \sys might employ various instrumentation techniques, necessitating multiple compilations of the target program, each with a distinct instrumentation pass. 
This process results in greater compilation overhead compared to using individual fuzzers. 
In our future research efforts, we plan to investigate more efficient instrumentation strategies to mitigate the overhead associated with diverse instrumentation requirements.

\noindent{\bf More sophisticated targets.}
The current version of \sys is optimized for the C/C++ programs in the Linux user space.
Our future work will extend \sys to other platforms and fuzzing domains, such as kernel fuzzing~\cite{googlesyzkaller,wang2021syzvegas,kim2020hfl} and compiler fuzzing~\cite{chen2020survey,li2022alphaprog,cummins2018compiler,xu2020dsmith}.
This extension will potentially demonstrate the effectiveness of \sys's collaboration strategy across various software layers and development tools, broadening the impact of collaborative fuzzing in the software security landscape.
\section{Conclusion}
\label{sec:conclusion}

In this paper, we introduce \sys, a novel collaborative fuzzing framework powered by a customized bandits model. 
Different from the greedy method utilized by the state-of-the-art collaborative fuzzing framework, \autofz, \sys model the long-term impact of individual fuzzers, learning globally optimal collaborative strategies.
Together with the bandits-based resource allocation, we also propose a new metric for fuzzer evaluation, a real-time seed synchronization mechanism, as well as a set of implementation-wise optimizations.
Through extensive experiments, we first demonstrate \sys's superior effectiveness, stability, and adaptability across a variety of programs, compared to \autofz and widely used individual fuzzers. 
Moreover, we conduct a comprehensive ablation study and hyper-parameter sensitivity test to verify \sys's key designs and its insensitivity to the changes in hyper-parameters.
Finally, we demonstrate \sys's effectiveness in bug detection for real-world programs by analyzing the results of a recent fuzzing competition, where \sys won the first place. 
Through these findings, we safely conclude that collaborative fuzzing presents a promising solution for constructing generic fuzzers, and multi-arm bandits can be used to design effective collaborative fuzzing strategies.
We hope our work can inspire further advancements in collaborative fuzzing through multi-arm bandits and reinforcement learning, to push the boundaries of software fuzzing in real-world applications.  %
\newpage

\bibliographystyle{IEEEtran}
{
\footnotesize
\bibliography{bib}
}

\appendix
\section{Appendix}



\subsection{Example of customized multi-arm bandit algorithm and resource allocation}
\label{appendix:thompson_example}

We present a comprehensive explanation of our customized multi-arm bandits, accompanied by a straightforward example that demonstrates the shift in weight distribution.

Let's assume that we use the method in \autoref{subsec:tech_overview} to schedule three fuzzers throughout the fuzzing campaign: AFL, AFLFast and AFL++.
At the beginning of a new round, say round $t$, the weight distribution for each fuzzer are as follows:
$AFL \sim Beta(2, 2)$, $AFLFast \sim Beta(3, 4)$, $AFL++ \sim Beta(5, 4)$.
For each fuzzer, we sample a value $\theta$ from its weight distribution.
Assume we got $\theta_{AFL}=0.57$, $\theta_{AFLFast}=0.32$, $\theta_{AFL++}=0.81$.
Since AFL++ has the highest sampled value, it is selected for the round to execute the fuzzing task.
After it finishes the task, we need to evaluate its performance by collecting the seeds it generated in this round. Then, we use \autoref{alg:evaluate} to evaluate those seeds and get a reward value $r$, let's say $r=0.78$.
We then need to discretize this value by sampling from a bernoulli distribution $Bernoulli(p=0.78)$, and use the sampled value as the discretized reward.
Assume we got $1$, then AFL++ obtains a reward equal to 1.
The reward distribution of AFL++ is then updated from $Beta(5,4)$ to $Beta(6, 4)$.

\begin{figure*}[ht]
    \centering
\includegraphics[width=1\linewidth]{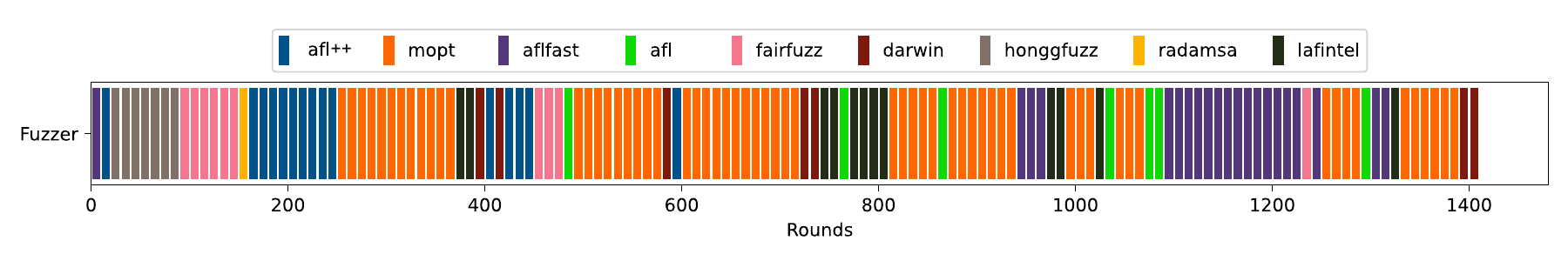}
    \caption{Bar graph representation of fuzzers achieving the highest mathematical expectation for reward distribution over consecutive rounds, sampled every 10 rounds, in a single trial targeting \texttt{lcms\_cms\_transform\_fuzzer}. Each colored bar indicates a fuzzer with the leading expectation value at that specific round.}
    \label{fig:weight_analysis}
\end{figure*}

The procedure offers the advantage of achieving a balance between exploration and exploitation. In real world scenarios, scheduling among fuzzers can be highly dynamic.
For instance, \autoref{fig:weight_analysis} presents a graphical representation of which fuzzer attains the maximum mathematical expected sampling value in each round. As depicted in the figure, no single fuzzer consistently outperforms others. AFL++, MOpt, and AFLFast each take turns leading in performance.

\begin{figure*}[t]
    \centering
    \includegraphics[width=1.02\linewidth]{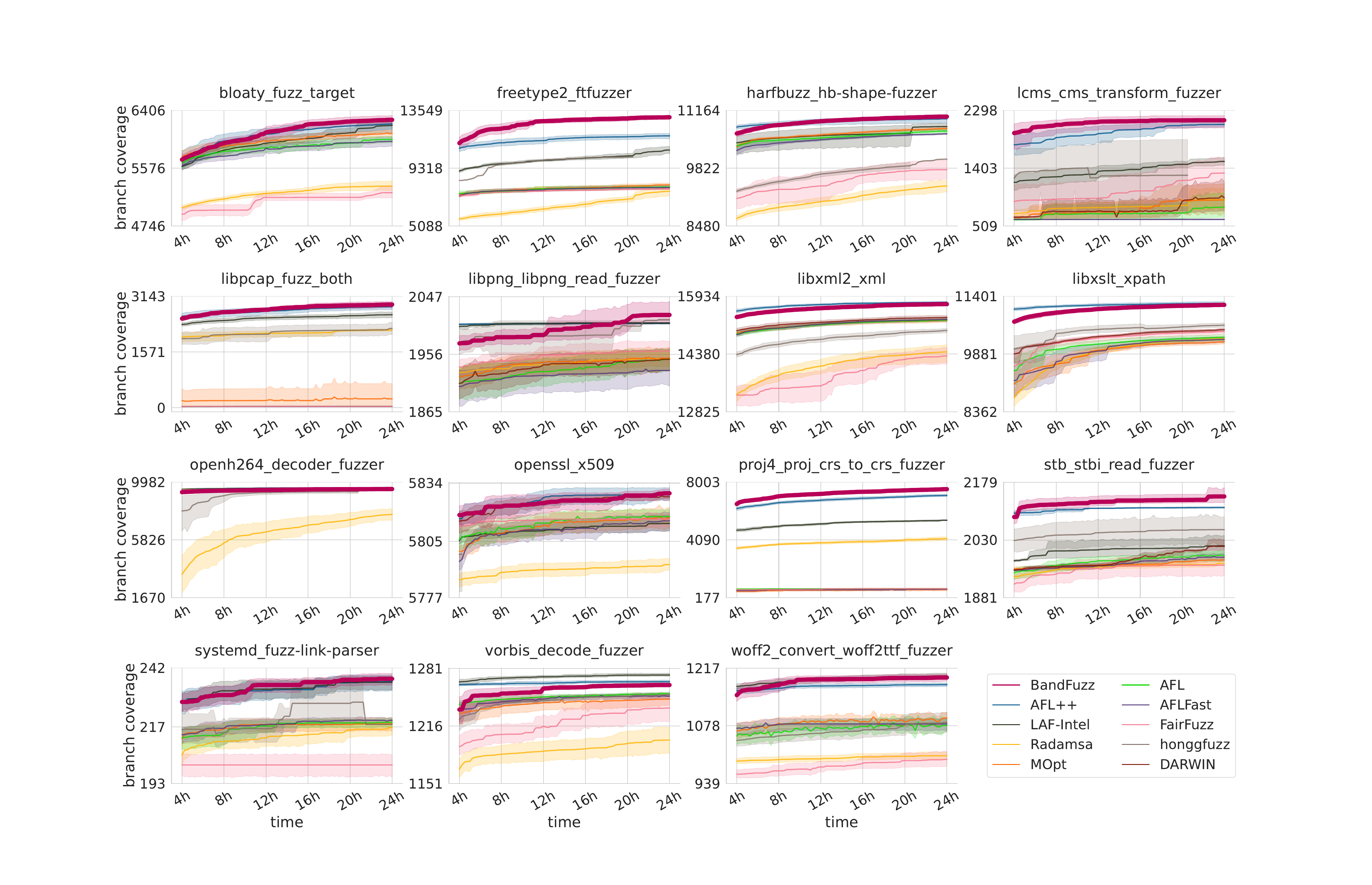}
    \caption{Evaluation of \sys and individual fuzzers on FuzzBench: Each line plot in the graph represents the arithmetic mean along with a 95\% confidence interval, derived from 10 independent trials. ``Branch coverage'' denotes the count of branches that each fuzzer has explored.}
    \label{fig:compare_individual}
\end{figure*}

\subsection{More FuzzBench Results}
\label{appendix:fuzzbench}

In \autoref{fig:compare_individual}, we show the branch coverage of \sys vs individual fuzzers across the time. 
The results are consistent with~\autoref{tab:fuzzbench}.

\end{document}